\documentclass[prl,aps,showpacs,groupedaddress,superscriptaddress,twocolumn,toc=flat]{revtex4-1}

\usepackage[table]{xcolor}

\usepackage[colorlinks=true,linkcolor=blue,urlcolor=blue,citecolor=blue]{hyperref}

\usepackage[utf8x]{inputenc}
\usepackage{color}
\usepackage{bbm} 

\usepackage{amsfonts,amsmath,amssymb,stmaryrd}

\usepackage{graphicx}
\usepackage{subfigure}  % use for side-by-side figures
\usepackage{bbm} 
\usepackage{hyperref}
\usepackage{epsfig}
\usepackage{mathrsfs}
\usepackage{verbatim}
\usepackage{centernot}
\usepackage{ulem}
\usepackage{longtable}
\usepackage{nicefrac}
\usepackage[framemethod=tikz]{mdframed}
 
\usepackage{booktabs}

\renewcommand{\l}{\left(}
\renewcommand{\r}{\right)}

\newcommand{\bra}[1]{\langle#1|}
\newcommand{\ket}[1]{|#1\rangle}
\newcommand{\bkt}[2]{\left\langle #1 |#2 \right\rangle}
\renewcommand{\ij}{{\langle \vec{i}, \vec{j} \rangle}}
\newcommand{\xy}{{\langle \vec{y}, \vec{x} \rangle}}

\renewcommand{\H}{\hat{\mathcal{H}}}

\renewcommand{\c}{\hat{c}}
\renewcommand{\a}{\hat{a}}

\newcommand{\f}{\hat{f}}
\newcommand{\cd}{\hat{c}^\dagger}

\newcommand{\ad}{\hat{a}^\dagger}

\newcommand{\hd}{\hat{h}^\dagger}
\newcommand{\h}{\hat{h}}

\newcommand{\hc}{\text{h.c.}}

\usepackage{array}

\usepackage{cancel,ifthen}
\newcommand{\cmnt}[2][NoInPuT]{\ifthenelse{\equal{#1}{NoInPuT}}{}{{\color{red}\sout{#1}}} {\color{blue} #2}}

\usepackage{bm}	% bold in math mode
\renewcommand{\vec}[1]{\bm{#1}}

\LTcapwidth=\textwidth

\begin{document}
\normalem	% changes \emph back to normal after introducing ulem package.

\title{Strong pairing in mixed dimensional bilayer antiferromagnetic Mott insulators}

\author{Annabelle Bohrdt}
\affiliation{Department of Physics and Institute for Advanced Study, Technical University of Munich, 85748 Garching, Germany}
\affiliation{Munich Center for Quantum Science and Technology (MCQST), Schellingstr. 4, D-80799 M\"unchen, Germany}
\address{ITAMP, Harvard-Smithsonian Center for Astrophysics, Cambridge, MA 02138, USA}
\affiliation{Department of Physics, Harvard University, Cambridge, MA 02138, USA}

\author{Lukas Homeier}
\affiliation{Department of Physics and Arnold Sommerfeld Center for Theoretical Physics (ASC), Ludwig-Maximilians-Universit\"at M\"unchen, Theresienstr. 37, M\"unchen D-80333, Germany}
\affiliation{Munich Center for Quantum Science and Technology (MCQST), Schellingstr. 4, D-80799 M\"unchen, Germany}

\author{Immanuel Bloch}
\affiliation{Max-Planck-Institut f\"ur Quantenoptik, 85748 Garching, Germany}
\affiliation{Department of Physics and Arnold Sommerfeld Center for Theoretical Physics (ASC), Ludwig-Maximilians-Universit\"at M\"unchen, Theresienstr. 37, M\"unchen D-80333, Germany}
\affiliation{Munich Center for Quantum Science and Technology (MCQST), Schellingstr. 4, D-80799 M\"unchen, Germany}

\author{Eugene Demler}
\affiliation{Department of Physics, Harvard University, Cambridge, MA 02138, USA}
\affiliation{Institute for Theoretical Physics, ETH Zurich, 8093 Zurich, Switzerland}

\author{Fabian Grusdt}
\email[Corresponding author email: ]{fabian.grusdt@physik.uni-muenchen.de}
\affiliation{Department of Physics and Arnold Sommerfeld Center for Theoretical Physics (ASC), Ludwig-Maximilians-Universit\"at M\"unchen, Theresienstr. 37, M\"unchen D-80333, Germany}
\affiliation{Munich Center for Quantum Science and Technology (MCQST), Schellingstr. 4, D-80799 M\"unchen, Germany}


\date{\today}

\begin{abstract}
Interacting many-body systems combining confined and extended dimensions, such as ladders and few layer systems are characterized by enhanced quantum fluctuations, which often result in interesting collective properties.
Recently two-dimensional bilayer systems, such as twisted bilayer graphene or ultracold atoms, have sparked a lot of interest because they can host rich phase diagrams, including unconventional superconductivity. Here we present a theoretical proposal for realizing high temperature pairing of fermions in a class of bilayer Hubbard models. We introduce a general, highly efficient pairing mechanism for mobile dopants in antiferromagnetic Mott insulators, which leads to binding energies proportional to $t^{1/3}$, where $t$ is the hopping amplitude of the charge carriers. The pairing is caused by the energy that one charge gains when retracing a string of frustrated bonds created by another charge. Concretely, we show that this mechanism leads to the formation of highly mobile, but tightly bound pairs in the case of mixed-dimensional Fermi-Hubbard bilayer systems. This setting is closely related to the Fermi-Hubbard model believed to capture the physics of copper oxides, and can be realized by currently available ultracold atom experiments. 
\end{abstract}

\maketitle

%%%%%%%%%%%%%%%%%%%%%%%%%%%%%%%%%%%%%%
\textbf{Introduction.--}
One of the fundamental results in the theory  of strongly interacting electron systems is the Kohn-Luttinger theorem, stating that electron pairing is inevitable in systems with purely repulsive interactions. 
Following the original Kohn-Luttinger paper in 1965 \cite{Kohn1965}, it was commonly assumed that their pairing mechanism gives rise to critical temperatures that are too small to be observable in realistic experimental systems. 
The discovery of high temperature superconductivity in cuprates \cite{Bednorz1986} brought renewed interest to the study of pairing in systems with repulsive interactions. 
Even today, one key question in the field of cuprate superconductors is whether electron pairing in these materials can be explained by purely repulsive Coulomb interactions. % or also involves phonon mediated attraction. 
The lack of theoretical consensus on this question motivated the development of the field of quantum simulation using ultracold atoms: 
The Fermi Hubbard model, commonly believed to capture the main properties of cuprate superconductors, is naturally realized using fermions in optical lattices \cite{Bloch2008,Tarruell2018,Bohrdt2021PWA}. 
During the last few years considerable experimental progress has been achieved in analyzing properties of this model including the observation of long-range antiferromagnetic order \cite{Mazurenko2017}, magnetic polaron properties \cite{Koepsell2019}, the crossover from polaron to Fermi liquid \cite{Koepsell2020_FL}, and bad metallic transport \cite{Brown2019a}.
However, the demonstration of pairing has so far been out of reach since temperatures significantly below the superexchange would be required. 

%%%%%%%%%%%%%%%%%%%%%%%%%%%%%%%%%%%%%%%%%%%%%%%%%%%%%
\begin{figure*}[t!]
\centering
  \includegraphics[width=0.95\linewidth]{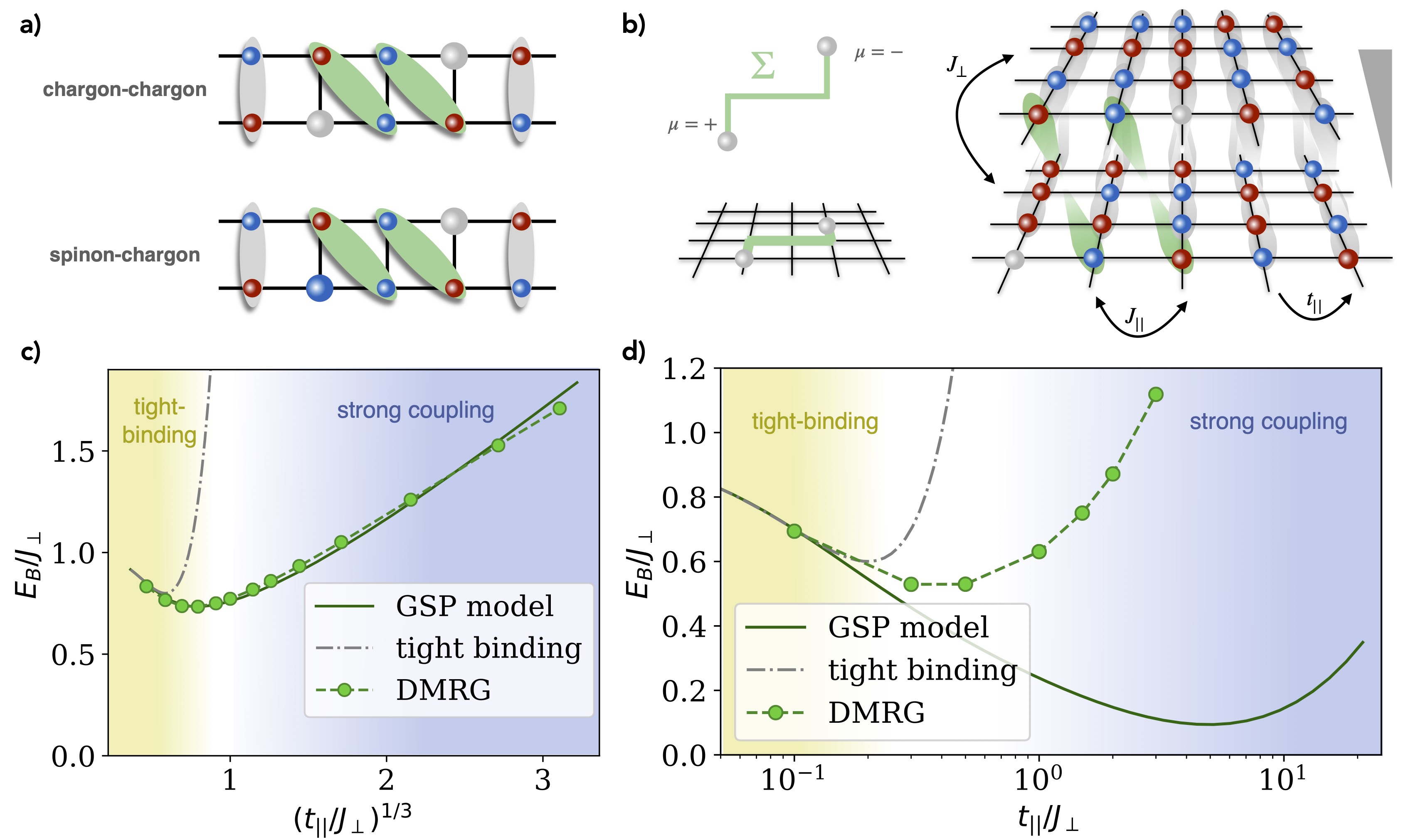}
\caption{\textbf{String-based chargon pairing in mixD bilayers.} We introduce an efficient pairing mechanism for distinguishable chargons with flavors $\mu=\pm$ connected by a string $\Sigma$. The string pairing mechanism can be realized e.g. in a model of $d$-dimensional bilayers a), b) with spin-$1/2$ particles and strong repulsive interactions. Concretely, we consider mixD systems, where the hopping $t_\perp$ between the layers is suppressed by a gradient $\Delta$, while the superexchange $J_\perp$ is kept intact, and $J_\perp \gg J_{||}$. The strong interlayer superexchange leads to the formation of rung singlets (dimers), depicted here for a) $d=1$ and b) $d=2$. The motion of doped holes via $t_\parallel$ within different layers $\mu=\pm$ tilts the singlets along their path, which corresponds to the formation of the string $\Sigma$. The potential energy associated with the string increases with its length as more singlets are tilted. Here, gray (green) shading corresponds to vertical (tilted) singlets. c), d) Binding energies $E_{\rm B}$ are obtained by comparing the energies of spinon-chargon and chargon-chargon mesons in the Gram-Schmidt parton model, and the ground state energy for a system with one versus two holes in DMRG \cite{Hauschild2018} as described in the text. We assume $J_{\parallel} / J_\perp=0.01$ and vary $t_{\parallel}/ J_\perp$ in Eq.~\eqref{eqeqHstring}. In the DMRG, weak inter-layer tunneling of strength $t_\perp/J_\perp=0.01$ was added to ensure convergence. In c), the resulting binding energies are shown for a $40\times 2$ ladder ($d=1$), and in d) for a bilayer system on a $12\times 4$  cylinder. 
}
\label{fig1}
\end{figure*}
%%%%%%%%%%%%%%%%%%%%%%%%%%%%%%%%%%%%%%%%%%%%%%%%%%%%%

The goal of this paper is twofold: on one hand, we identify a microscopic bilayer model that features repulsive interactions between fermions, yet exhibits pairing at temperatures comparable to the superexchange, realizable by state of the art quantum simulators \cite{Bohrdt2021PWA}, see Fig.~\ref{fig1}. 
A special characteristic of this model is its mixed dimensional (mixD) character: it features interlayer exchange interactions, but no interlayer single particle tunneling \cite{Grusdt2018mixD,Grusdt2020}, Fig.~\ref{fig1}a),b). Such systems can be realized experimentally using ultracold atoms by applying a potential gradient in bilayer and ladder models,  which have been recently realized in cold atom experiments \cite{Koepsell2020,Hartke2020,Gall2021,Sompet2021}. 

The second goal of our paper is to reveal a general mechanism for pairing in doped antiferromagnetic Mott insulators (AFMI). It is based on the idea that a single hole can be understood as a bound state of chargon and spinon \cite{Read1983,Coleman1984,Coleman1987,Wen1996,Vijayan2020}, carrying the respective charge and spin quantum numbers of the underlying AFMI \cite{Beran1996,Laughlin1997,Senthil2003,Bohrdt2020_ARPES,Bohrdt2021arXiv}, which are connected by a geometric string of displaced spins. The antiferromagentic correlations (which may also be short-ranged) of the parent state are assumed to induce a linear string tension \cite{Chiu2019Science}. 
Within the same framework, a bound state of two holes corresponds to a bound state of two chargons. We demonstrate that the interplay of potential energy, arising from the string tension, and kinetic energy of the partons leads to large binding energies for mesonic pairs of chargons. Our work goes beyond earlier analysis \cite{Bulaevskii1968,Trugman1988,Shraiman1988a,Manousakis2007,Vidmar2013} by including adverse effect that suppress pairing in a generic system. Effects detrimental to pairing will be suppressed in the mixed dimensional bilayer model we propose, featuring overwhelmingly attractive emerging parton interactions.

In the present work we focus on the case without interlayer tunneling, $t_\perp=0$, which enables an effective theoretical description. We consider strong spin-charge coupling, where the intralayer tunneling $t_{||} \gg J_\perp$ exceeds the relevant superexchange $J_\perp$ (see Fig.~\ref{fig1}). We demonstrate that the deeply bound pairs are highly mobile. This constitutes a notable exception from the common expectation \cite{Kivelson2020_talk} that bipolarons should carry a heavy effective mass. The light mass of bipolarons is also significant, because it allows for high condensation temperatures $T_c$ and has important implications for the ground state at finite doping: a light mass renders the localization of pairs unlikely. Therefore, we expect density wave states, such as stripes, to be less competitive with the superfluid states of pairs. In the opposite limit, $t_{||} \ll J_\perp$, we also expect pairing, but in this case the pairs are heavy and thus $T_c$ is low \cite{Bohrdt2021PWA}.

The microscopic mixD bilayer model we study hosts a BEC regime with mobile but tightly bound pairs at low doping, and a BCS regime at high doping \cite{Bohrdt2021PWA}.
Hence our findings may have further implications for understanding the phenomenology of high-temperature superconductivity: When  the interlayer couplings, denoted $J_\perp$, are switched off, the bilayer systems we consider reduce to independent copies of the two-dimensional Fermi-Hubbard model. The latter is widely believed to capture the essential physics of the cuprate superconductors. While experiments indicate that the cuprates do not realize a BEC phase with tightly bound Cooper pairs, they may be close in parameter space to models which exhibit a full BEC-to-BCS cross-over. This would help explain the pseudogap observed in cuprates, which is reminiscent of the pseudogap that arises at strong coupling in a BEC-to-BCS crossover, as realized in the mixD bilayer model.

%%%%%%%%%%%%%%%%%%%%%%%%%%%%%%%%%%%%%%
\textbf{String-based chargon pairing.--}
%%%%%%%%%%%%%%%%%%
The main result of this article is the identification of a general string-based pairing mechanism for doped holes in AFMIs, see Fig~\ref{fig1}. 
We begin by introducing  a general formulation of the model, which we dub the dimer-parton model, in an abstract way not tied to any microscopic Hamiltonian. We present assumptions that go into the construction of the model and summarize its key predictions, and delegate technical details to the Methods section. 
 In the following section we propose a microscopic bilayer model, see Fig.~\ref{fig1}a),b), and demonstrate that it realizes the abstract string scenario we envision. In particular, this allows to understand the surprisingly high binding energies we find for mobile, i.e. strongly coupled, holes in two-dimensional mixD bilayers shown in Fig.~\ref{fig1}d). 

In our general dimer parton model we describe excitations of AFMIs by two different types of partons: spinons and chargons, which carry the spin and charge quantum number, respectively. The partons also carry a flavor index $\mu=\pm$, allowing us to work with distinguishable chargons without a mutual hard-core constraint, which we predicted to pair up at strong couplings. The two different flavors $\mu=\pm$ could, for example, be internal degrees of freedom or different layers in a bilayer system, see Fig.~\ref{fig1}. 
In the dimer parton model we can treat arbitrary lattice geometries, but we assume a homogeneous coordination number $z$ for all sites; e.g. $z=4$ for the bilayer system shown in Fig.~\ref{fig1}b).

Furthermore, we consider rigid strings $\Sigma$ on the underlying lattice, which connect the partons and fluctuate only through the motion of the latter. We assume a linear string tension, i.e. a string is associated with an energy cost $V(\ell_\Sigma) = \sigma_0 \ell_\Sigma$ proportional to its length $\ell_\Sigma$, which reflects how the short-range antiferromagnetic correlations of the underlying AFMI are disturbed by the parton motion. In the microscopic mixD bilayer models discussed below, see Fig.~\ref{fig1}a),b), this string tension corresponds to the cost of breaking up rung singlets by the parton motion. 

Since we consider rigid strings with a linear string tension, the partons are always confined in our model, forming mesonic states.
The binding energy for two charges is hence obtained by comparing the ground state energy of a spinon-chargon (sc) pair to the energy per chargon of a chargon-chargon (cc) pair,
\begin{equation}
 	E_{\rm B} = 2 E_{\rm sc} - E_{\rm cc} .
	\label{eqEBdef}
\end{equation} 
At strong couplings, i.e. when the hopping amplitude $t \gg \sigma_0$ of the holes exceeds the string tension $\sigma_0 \propto J$ determined by the superexchange $J$, we predict a  binding energy with a characteristic scaling 
\begin{equation}
 	E_{\rm B} =  \alpha ~ \underbrace{(2 - 2^{1/3})}_{=0.740...} ~ t^{1/3} \sigma_0^{2/3} + \mathcal{O}(J),
 	\label{eqEBstring}
\end{equation}
where $\alpha > 0$ is a non-universal constant proportional to $(z-1)^{1/6}$; see Methods for a detailed derivation. 

Since a combination of $t$ and $\sigma_0 \propto J$ appears in Eq.~\eqref{eqEBstring}, we predict a remarkably strong asymptotic binding energy. The appearance of $t$ in this expression highlights the key aspect of the underlying binding mechanism, where two chargons share one string and gain equal amounts of superexchange and kinetic energy: The motion of one chargon frustrates the underlying AFMI along the string $\Sigma$. In the case of two chargons, the second chargon can retrace the string created by the first, thus lowering the energy cost due to frustration in the spin sector. $E_{\rm B}$ becomes positive because binding to a light second chargon with mass $\propto 1/t$ is favorable compared to binding two chargons individually to heavy spinons with mass $\propto 1/J$.

In a 2D square lattice $\alpha \approx 2$ and using a string tension of $\sigma_0 \approx J = 4 t^2/U$ we expect a binding energy on the order of $|E_{\rm B}| \simeq t^{1/3} J^{2/3} \simeq 2.5 t (t/U)^{2/3}$. This suggests that binding energies a significant fraction of a typical tunnel coupling $t$ may be possible, at least in principle, potentially exceeding room temperature. 

As a second main result of the dimer parton model we estimate the effective mass of the chargon pair at strong couplings,
\begin{equation}
    M_{\rm cc}^{-1} = 4 t \sqrt{z-1} / z.
    \label{eqMccMain}
\end{equation}
Despite being tightly bound, the pair is highly mobile -- contrary to common expectations for bipolarons.

We anticipate that the general string-based pairing mechanism plays a role in different microscopic models of AFMIs. Below, we discuss a specific realization in the mixD bilayer systems shown in Fig.~\ref{fig1}. 

%%%%%%%%%%%%%%%%%%%%%%%%%%%%%%%%%%%%%%
\textbf{Microscopic model.--}
%%%%%%%%%%%%%%%%%%
We consider a bilayer of $d$-dimensional sheets with spin-$1/2$ particles $\c_{\vec{j},\mu,\sigma}$, which can be either bosons or fermions. Here, $\mu,\sigma=\pm$ are layer and spin indices, respectively. In the following, we consider fermions for concreteness. The particles are assumed to be strongly repulsive, allowing us to work in a subspace without double occupancies. The Hamiltonian we consider includes NN hopping $t$ within the layers and spin-exchange couplings $J_{\mu,\mu'}=J_{\mu',\mu}$:
\begin{multline}
 	\H = - t_\parallel~  \hat{\mathcal{P}} \sum_{\ij} \sum_{\mu, \sigma=\pm} \l \cd_{\vec{i},\mu,\sigma} \c_{\vec{j},\mu,\sigma} + \hc \r \hat{\mathcal{P}} +\\
	+ \sum_{\langle \vec{i}_\mu \vec{j}_{\mu'}\rangle} J_{\mu,\mu'} \l  \hat{\vec{S}}_{\vec{i}\mu} \cdot \hat{\vec{S}}_{\vec{j},\mu'}  - \frac{1}{4} \hat{n}_{\vec{i},\mu} \hat{n}_{\vec{j},\mu'} \r,
\label{eqeqHstring}
\end{multline}
where $\hat{\mathcal{P}}$ projects to the subspace with maximum single occupancy per site; $\hat{\vec{S}}_{\vec{j},\mu}$ and $\hat{n}_{\vec{j},\mu}$ denote the on-site spin and density operators, respectively. The sum in the second line includes NN bonds within ($\mu=\mu'$) and between ($\vec{i}=\vec{j}$) the layers, and we set $J_{++}=J_{--}=J_{||}$ and $J_{+-}=J_\perp>0$, see Fig.~\ref{fig1}b).

Experimentally, the model in Eq.~\eqref{eqeqHstring} can be realized in $d=1,2$ starting from a Hubbard Hamiltonian \cite{Bohrdt2021PWA} with on-site interactions $U$. A strong inter-layer gradient $\Delta$ can be used to obtain a mixD setting \cite{Grusdt2018mixD} along the third direction where the two layers $\mu=\pm$ are physically realized. Note that the AFM coupling $J_\perp > 0$ can be realized both for fermions and bosons by an appropriate choice of the gradient $\Delta$ and the Hubbard interaction $U$ \cite{Duan2003,Dimitrova2019,Sun2020}. In the following we assume that the inter-layer spin-exchange is dominant, $J_\perp \gg |J_\parallel|$, which can be achieved by choosing $U \gg t_{||}$.

As a central result of our article, we predict strong binding energies of holes, on the order of $|E_{\rm B}| \gtrsim J_\perp$, in the strong coupling regime $t_\parallel \gg J_\perp$ where charges are highly mobile. In Fig.~\ref{fig1}c),d) we show binding energies obtained from DMRG simulations in mixD ladders ($d=1$) and bilayers ($d=2$) for $J_\parallel=0$. The strong binding observed numerically can be explained by the string-based pairing mechanism introduced above. 

The undoped ground state of Eq.~\eqref{eqeqHstring} at half filling and for $|J_\parallel| \ll J_\perp$ corresponds to a product of rung singlets, $\ket{\Psi_0} = 2^{-V/2} \prod_{\vec{j}} ( \ket{ \!\! \uparrow \downarrow}_{\vec{j}} - \ket{\! \! \downarrow \uparrow}_{\vec{j}} )$, where $V=L^d$ is the volume. Excitations correspond to breaking up singlets and doping the system with holes. The motion of a hole introduces tilted singlets along its path, directly imprinting the string $\Sigma$ between two partons into the spin background. As illustrated in Fig.~\ref{fig1}a),b), the mixD bilayer model naturally realizes  spinon-chargon and chargon-chargon bound states.

%%%%%%%%%%%%%%%%%%%%%%%%%%%%%%%%%%%%%%
\textbf{Effective parton models.--}
%%%%%%%%%%%%%%%%%%
In the following, we discuss how the microscopic model, Eq.~\eqref{eqeqHstring} can be related to effective parton models, see \cite{SI} for a detailed derivation. We consider systems with two distinguishable partons, $n=1,2$, where the label $n$ summarizes the set of properties layer $\mu$, parton type (spinon or chargon), and spin $\sigma$. In order to obtain distinguishable partons, at least one of these properties has to be different between $n=1$ and $n=2$. 
We introduce the notation $\ket{\vec{x}_1,\vec{x}_2,\Sigma}$, where $\Sigma$ is a string of tilted singlets connecting $\vec{x}_2$ to $\vec{x}_1$, and $\vec{x}_n$ are the positions of the partons. 

Below, we discuss two parton models, the dimer parton model and the refined Gram-Schmidt parton (GSP) model. They can be related to the model in Eq.~\eqref{eqeqHstring} within the following approximations \cite{SI}:
\begin{itemize}
    \item[(i)] The frozen-spin approximation \cite{Grusdt2019} neglects fluctuations of spins along the string $\Sigma$; this is justified by a separation of time scales when $t_\parallel \gg J_\perp$.
    \item[(ii)] We neglect effects of self-crossings of strings, which becomes exact for large coordination numbers $z\gg 1$ (or in ladders with $d=1$). Even for a $d=2$ square lattice bilayer, such effects are expected to be quantitatively small due to their small relative share of the string Hilbertspace \cite{Grusdt2018tJz}. 
    \item[(iii)] In the dimer parton model, following the spirit of the Rokhsar-Kivelson quantum dimer model \cite{Rokhsar1988}, we assume that spin configurations $\ket{\vec{x}_1,\vec{x}_2,\Sigma}$ with parton positions $\vec{x}_{1,2}$ form an approximately orthonormal basis, $\bra{\vec{x}_1',\vec{x}_2',\Sigma'} \vec{x}_1,\vec{x}_2,\Sigma \rangle \approx \delta_{\vec{x_2}',\vec{x}_2} \delta_{\vec{x_1}',\vec{x}_1} \delta_{\Sigma',\Sigma}$, which is justified by the small overlaps of configurations with non-identical singlet configurations. This leads to an effective dimer parton Hamiltonian with nearest-neighbor hopping of the partons under simultaneous adaption of the string states, and a string potential $V_\Sigma$, see Methods for details. 
\end{itemize}

The third approximation differs between the dimer parton and the GSP model. 
In the GSP model, the assumption that string states form an orthonormal basis is partially dropped. To this end we perform a Gram-Schmidt orthogonalization of the string states defined in the $t-J$ Hilbert space and account for the non-orthogonal nature of sc states, up to loop effects \cite{SI}. This leads to an accurate prediction of the spinon dispersion relation. For chargon-chargon states, the GSP and dimer parton models are equivalent.

In the parton models, for simplicity, we assume a linear string potential,
$V^{\rm sc}_\Sigma = \sigma_0 ~ \ell_\Sigma$, where $\ell_\Sigma$ is the length of string $\Sigma$. For the tilted singlets in the mixD bilayer setting, Eq.~\eqref{eqeqHstring}, the linear string tension is $\sigma_0 =\nicefrac{3}{4}  J_\perp$, corresponding to the energy difference between a singlet and a random spin configuration. In the chargon-chargon case, the string potential has an additional on-site attraction, $V_\Sigma^{\rm cc} = V_\Sigma^{\rm sc} -\delta_{\Sigma,0} J_\perp/4$, which leads to slightly stronger binding than predicted by Eq.~\eqref{eqEBstring}.

Two holes are distinguishable if they move within separate layers $\mu_1 = - \mu_2$, and they can thus occupy the same position $\vec{x}$ in the lattice. Note that the motion of a second hole in the opposite layer and along the same path can completely remove the geometric string $\Sigma$ of displaced spins. 

%%%%%%%%%%%%%%%%%%%%%%%%%%%%%%%%%%%%%%
\textbf{Numerical results.--}
%%%%%%%%%%%%%%%%%%
In Fig.~\ref{fig1}c) we show the binding energies predicted by the GSP model for the mixD ladder ($d=1$). Plotting the results over $(t_\parallel / J_\perp)^{1/3}$ shows that the binding energy at strong couplings, $t_\parallel > J_\perp$, scales as $|E_{\rm B}| \simeq t_\parallel^{1/3} J_\perp^{2/3}$ as predicted by the dimer parton model. In the opposite tight-binding regime, $t_\parallel \ll J_\perp$, we find excellent agreement with our perturbative result from Ref.~\cite{Bohrdt2021PWA}, see dash-dotted line in Fig.~\ref{fig1}c).
For our DMRG simulations, we define the binding energy as $E_B = 2E_1 - (E_0+E_2)$, where $E_N$ is the ground state energy of the state with $N$ holes. Our DMRG results are in excellent agreement with our prediction by the GSP model for couplings $t_\parallel / J_\perp$ ranging over several orders of magnitude, from $0.1$ to $>30$.

%%%%%%%%%%%%%%%%%%%%%%%%%%%%%%%%%%%%%%%%%%%%%%%%%%%%%
\begin{figure}[t!]
\centering
  \includegraphics[width=0.92\linewidth]{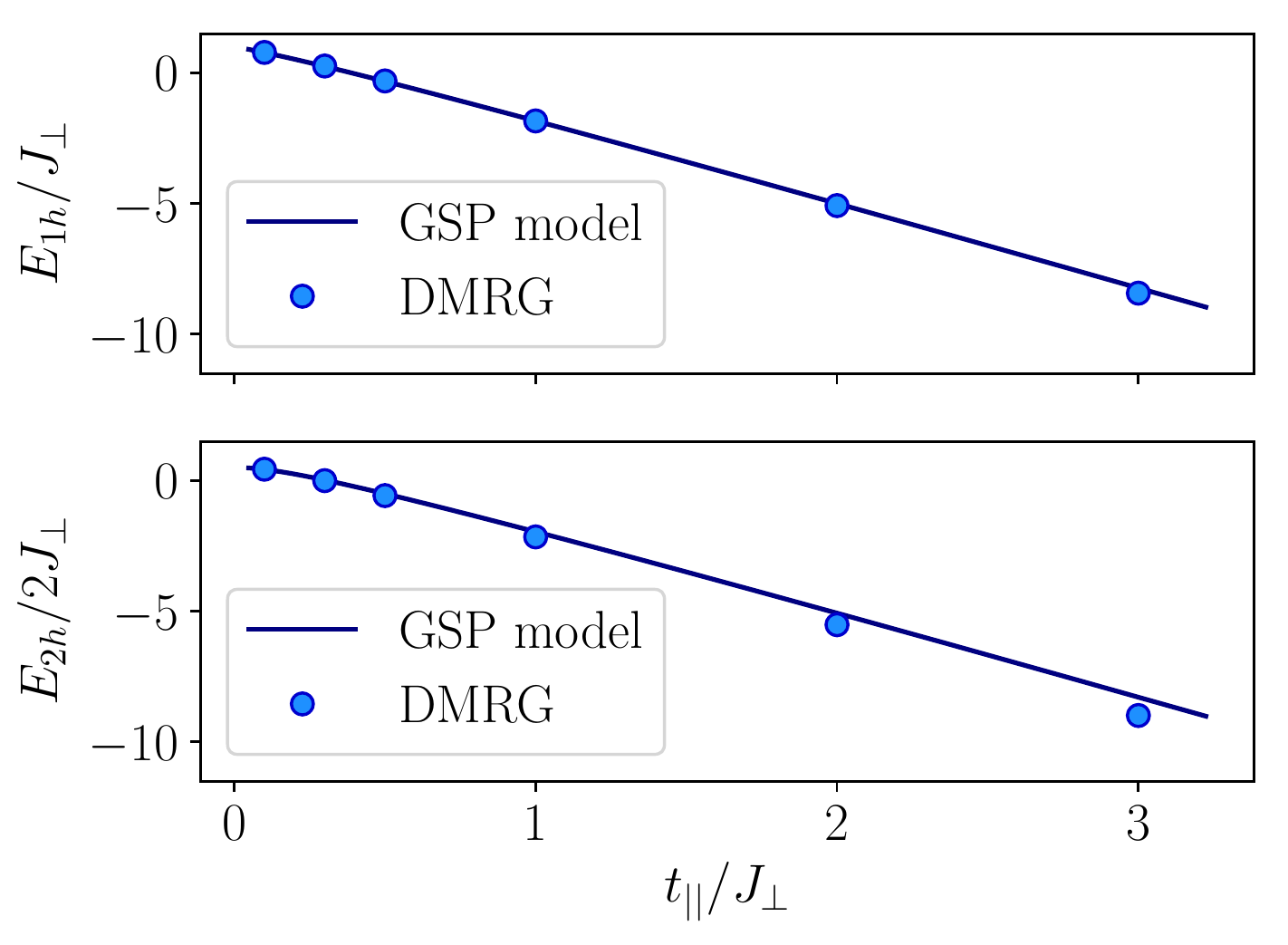}
\caption{\textbf{MixD bilayer in 2D.} We study ground state energies in the mixD bilayer system with $d=2$ using the Gram-Schmidt parton (GPS) model and DMRG for a bilayer system on a $12\times 4$ cylinder. We show the energy per hole as a function of $t_{||}/J_\perp$ for one (top) and two holes (bottom).} 
\label{fig:2D}
\end{figure}
%%%%%%%%%%%%%%%%%%%%%%%%%%%%%%%%%%%%%%%%%%%%%%%%%%%%%

In Fig.~\ref{fig1}d), we show our results for the mixD bilayer system ($d=2$). From the dimer parton model we again expect strong pairing when $t_\parallel \gg J_\perp$. This prediction is confirmed by our numerical DMRG results in $2 \times 12 \times 4$ systems, where the first direction denotes the two layer indices and we assume periodic (open) boundary conditions along the short (long) axes of the cylinder. For $t_\parallel/J_\perp = 3$ we find remarkably strong binding energies below the superexchange scale $J_\perp$. For small $t_\parallel/J_\perp \ll 1$ the perturbative tight-binding result \cite{Bohrdt2021PWA} is obtained.

We also compare predictions by the GSP model to our DMRG results for $d=2$. In Fig.~\ref{fig:2D}a) we show the energies per hole in the spinon-chargon and chargon-chargon case, respectively, and obtain very good agreement. However, a small deviation of a few percent can be observed at strong couplings in the two-hole case, where the GSP and dimer parton models are identical. We expect that loop effects, ignored in the parton models, lead to the slightly lower energy per hole found by DMRG. Since the binding energy, shown in Fig.~\ref{fig1}d), is obtained as the small difference between the large one- and two-hole energies, see Eq.~\eqref{eqEBstring}, the deviation of the GSP model from DMRG appears sizable at strong couplings. We emphasize however that GSP and DMRG consistently predict positive binding energies throughout, and a cross-over from tightly bound to strongly fluctuating extended pairs of holes.

\begin{figure}[t!]
\centering
  \includegraphics[width=0.99\linewidth]{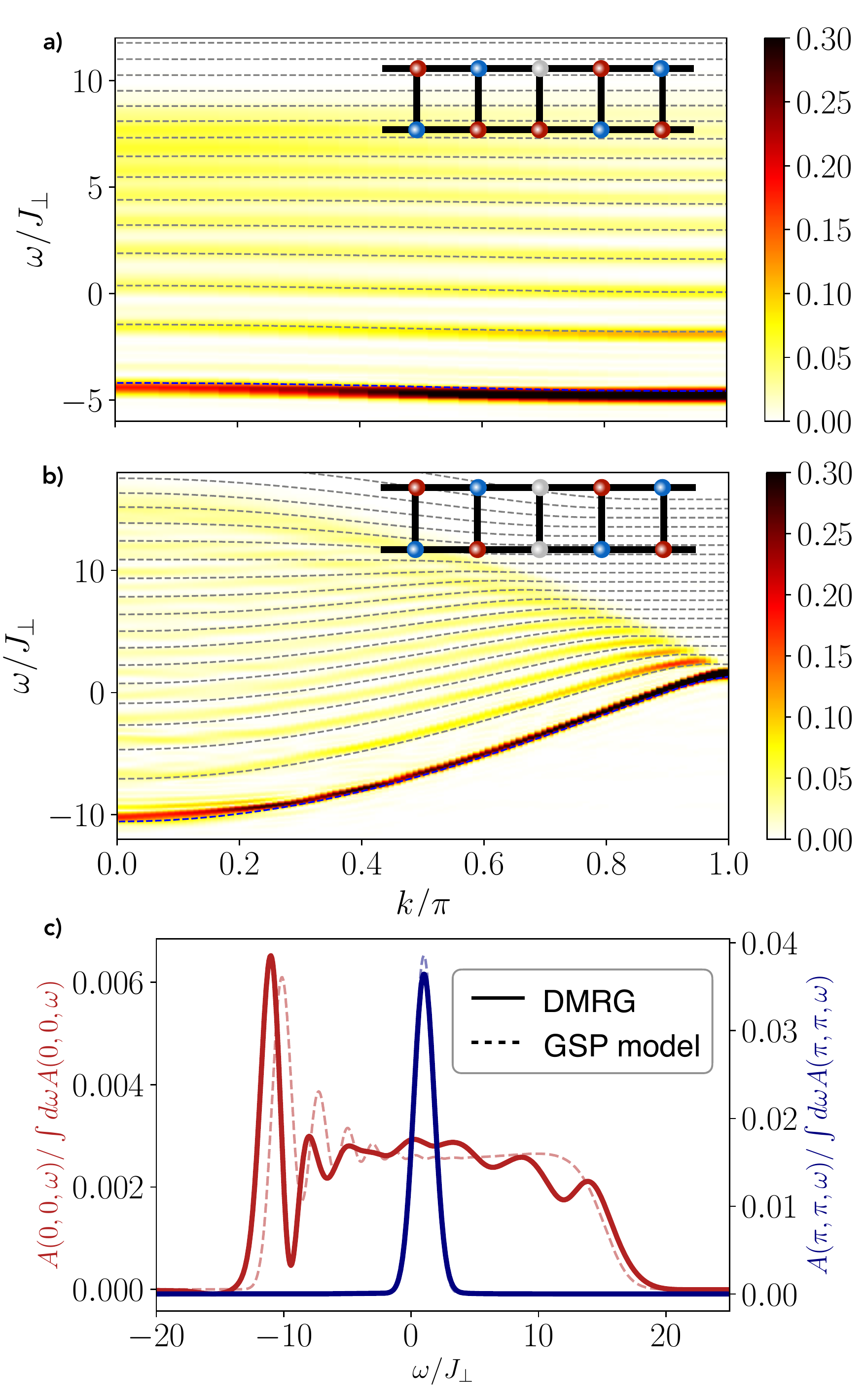}
\caption{\textbf{Spectral function in mixed dimensions.} Spectra for a) one hole and b) two holes in a ladder ($d=1$), and c) two holes in a mixD bilayer ($d=2$). In a) and b) we assume $t_\parallel/J_\perp = 3.4$ and $J_\parallel/J_\perp=0.31$; in c) $t_\parallel / J_\perp = 2$ and $J_\parallel/J_\perp = 0.01$. In our time-dependent DMRG (colormap) we added weak tunneling $t_\perp/J_\perp=0.01$ to simplify convergence. In a), b) the dashed lines correspond to eigenenergies of the GSP model \cite{SI}, where only states with even inversion symmetry are shown. The observed discrete lines correspond to vibrational excitations of emerging spinon-chargon and chargon-chargon mesons in the system. In c) the dashed lines describe GSP spectra broadened by the Fourier resolution of our time-dependent DMRG simulations (solid lines). Red (blue) lines correspond to momenta $\vec{k}=\vec{0}$ ($\vec{k}=\vec{\pi}$).} 
\label{fig:spectra}
\end{figure}

%%%%%%%%%%%%%%%%%%%%%%%%%%%%%%%%%%%%%%
\textbf{Signatures of parton-formation in one- and two-hole ARPES spectra.--}
%%%%%%%%%%%%%%%%%%%%%%%%%%%%%%%%%%%%
To further corroborate the parton structure of the one- and two-hole states we found at strong coupling and demonstrate the high degree of mobility of the paired states, we investigate their spectral properties. We performed time-dependent DMRG simulations \cite{Kjall2013,Zaletel2015,Paeckel2019} in the mixD ladder ($d=1$) and bilayer ($d=2$) to extract the following spectral functions,
\begin{equation}
    A_{1/2}(\vec{k},\omega) = \sum_n \delta(\omega-E^{(1/2)}_n+E_0)|\bra{\psi^{(1/2)}_n} \hat{C}_{(1/2),\vec{k}} \ket{\psi_0}|^2,
    \label{eqSpectralFunction}
\end{equation}
where $\ket{\psi_0}$ ($E_0$) is the ground state (energy) without holes, and $\ket{\psi_n^{(1/2)}}$ ($E^{(1/2)}_n$) are the eigenstates (energies) with one and two holes, respectively. 
The excitations are created through the operators $\hat{C}_{(1/2),\vec{k}}$, which are defined as the Fourier transforms of $\hat{C}_{1,\vec{i}} = \c_{\vec{i},+,\uparrow}$ and $\hat{C}_{2,\vec{i}} = \c_{\vec{i},+,\uparrow} \c_{\vec{i},-,\downarrow}$, see insets in Fig.~\ref{fig:spectra}a) and b).

%%%%%%%%%%%%%%%%%%%%%%%%%%%%%%%%%%%%%%%%%%%%%%%%%%%%%
%%%%%%%%%%%%%%%%%%%%%%%%%%%%%%%%%%%%%%%%%%%%%%%%%%%%%

Our results in Fig.~\ref{fig:spectra}a), b) for the ladder ($d=1$) reveal a series of narrow lines which are visible up to high energies, in both the one-hole a) and two-hole b) spectra. They correspond to long-lived vibrational excitations of the spinon-chargon and chargon-chargon meson, respectively. This intuition is confirmed by the excellent agreement we obtain with energies predicted by the GSP model. In \cite{SI} we show that the parton model also captures the spectral weights. The existence of discrete internal excitations, as revealed in Fig.~\ref{fig:spectra}, is a hallmark of emerging mesonic states in doped AFMIs. In addition to the vibrational states visible in Fig.~\ref{fig:spectra}, we also predict odd-parity spinon-chargon and chargon-chargon states \cite{Bohrdt2021arXiv} from the parton model, whose spectral weight is negligible in Fig.~\ref{fig:spectra}, see \cite{SI}.

For the two-hole case, see Fig.~\ref{fig:spectra}b), we observe a strong center-of-mass dispersion. This indicates that the deeply bound pair is highly mobile, confirming a key prediction of the parton model, see Eq.~\eqref{eqMccMain}. Additionally, at $k_x=\pi$ the relative motion of the partons is suppressed by quantum interference, and only the lowest mesonic state has non-vanishing spectral weight.

The parameter set we used in Fig.~\ref{fig:spectra}a), b) assumes $J_\parallel / J_\perp=0.31$ and corresponds to a situation which could be realistically realized in a Fermi-Hubbard ladder with a potential gradient between the two legs \cite{Bohrdt2021PWA}. Hence, the emerging spinon-chargon and chargon-chargon meson states we predict in our model should be experimentally accessible with ultracold atoms in optical lattices with current technology. 

In Fig.~\ref{fig:spectra}c) we show two-hole spectral cuts at rotationally invariant momenta $\vec{k}=\vec{0}=(0,0)$ and $\vec{\pi}=(\pi,\pi)$ for the mixD bilayer where $d=2$; we consider $t_\parallel/J_\perp=2$ in the strong coupling regime. Here, our underlying time-dependent DMRG simulations on a $2\times40\times4$ cylinder are more challenging and we reach shorter times than in $d=1$. This results in significant Fourier broadening, and prevents us from identifying individual vibrational excitations at high energies. Nevertheless, at $\vec{k}=\vec{0}$ a first peak is visible in the DMRG data. Comparison with a broadened GSP model spectrum suggests that this peak likely corresponds to the first vibrational meson state. 

At $\vec{k}=\vec{\pi}$, the two-hole spectrum in Fig.~\ref{fig:spectra}c) collapses to a single peak. As in $d=1$ at $k=\pi$ this can be understood within the dimer parton model from destructive quantum interference which leads to a cancellation of the chargon's relative motion. The energy of the tightly-bound two-hole state at $\vec{k}=\vec{\pi}$ is accurately predicted by the GSP model. 

As suggested by the large energy difference of the chargon-chargon state between $\vec{k}=0$ and $\vec{k}=\pi$, on the order of $6 t_\parallel$ in Fig.~\ref{fig:spectra}c), we find that the deeply bound state at $\vec{k}=\vec{0}$ has a light effective mass. A fit $k_x^2/(2 M_{\rm cc})$ to the lowest spectral peak $E_{\rm cc}(k_x \vec{e}_x) - E_{\rm cc}(\vec{0})$ yields $M_{\rm cc} = 2.44 /t_\parallel$ at $t_\parallel / J_\perp=2$. This compares reasonably with our estimate $M_{\rm cc} \geq 0.63 / t_\parallel$ from the dimer-parton model; we used an improved version of Eq.~\eqref{eqMccMain}, see \cite{SI}. This result should be contrasted with the much heavier $M_{\rm cc} = 25 / J_\perp$ in the tight-binding regime \cite{Bohrdt2021PWA}, at $t_\parallel / J_\perp = 0.1$ with a comparable binding energy of order $|E_{\rm B}| \approx \nicefrac{3}{4} J_\perp$.

%%%%%%%%%%%%%%%%%%%%%%%%%%%%%%%%%%%%%%
\textbf{Finite-temperature phase diagram.--}
%%%%%%%%%%%%%%%%%%
At finite temperatures, the emergence of two kinds of mesons can lead to rich physics. While the ground states we find are always paired, we expect a mixture of spinon-chargon and chargon-chargon pairs in chemical equilibrium when $T>0$. Because a gas of $N$ spinon-chargon pairs has more entropy than a gas of only $N/2$ chargon-chargon pairs at high temperatures, we expect a cross-over from spinon-chargon to chargon-chargon domination around some temperature $T_{\rm mix}$. 

In $d \geq 2$ dimensions, we expect an even richer phase diagram. The chargon-chargon pairs can form a (quasi-) condensate below ($T_{\rm BKT}$) $T_c$ in ($d=2$) $d > 2$ dimensions at finite doping and before sc pairs dominate beyond $T_{\rm mix}$. Furthermore, at a critical temperature $T^*$ on the order of $J_\perp$ we expect a thermal deconfinement transition of the mesons to take place when $d\geq 2$ \cite{BohrdtHahn2021prep}.

%%%%%%%%%%%%%%%%%%%%%%%%%%%%%%%%%%%%%%
\textbf{Discussion and outlook.--}
%%%%%%%%%%%%%%%%%%
We introduced a general string-based pairing mechanism for distinguishable chargons in AFMIs, and demonstrated that the resulting binding energy scales as $|E_{\rm B}| \simeq t^{1/3} J^{2/3}$, or $|E_{\rm B}| \simeq t (t/U)^{2/3}$, when $t \gg J$. The highest binding energies are expected when the chargon-chargon mesons become extended, with an average string length scaling as $\ell \simeq (t/J)^{1/3}$ \cite{Bulaevskii1968}, see supplementary  \cite{SI}. We confirmed the validity of our analytical arguments by numerical DMRG simulations of bilayer toy models which can be realized experimentally using e.g. ultracold atoms in optical lattices \cite{Bohrdt2021PWA}. Another interesting situation corresponds to cuprate materials with a multilayer structure and an effective chemical potential offset between the layers \cite{Chen2006,Shimizu2007}: Starting from half filling, we envision that a strong pump pulse may allow to create meta-stable pairs of doublons and holes in opposite layers. These may then form the mesonic bound states we propose, and could be probed by a subsequent laser pulse.

The analysis we performed here is limited to individual mesons, but our prediction of deeply bound and highly mobile chargon pairs for $t \gg J$ has important implications at finite doping. There we expect for the mixD bilayer ($d=2$) a quasi-condensate with power-law correlations below a critical BKT temperature $T_{\rm c}$. Owing to the high mobility of the chargon-chargon pairs, a sizable $T_{\rm c}$ at a fraction of $t$, and robustness against localization is expected.

Another interesting direction would be the investigation of an emergent $\rm{SO}(5)$ symmetry in the bilayer models, see Ref.~\cite{Scalapino1998,Demler2004} for a study in a related model.
%While the general pairing mechanism we propose should similarly be present for indistinguishable holes in the doped 2D Fermi-Hubbard model, several competing mechanisms need to be considered in that case: (i) indistinguishable chargons are mutually hard-core, which forbids occupation of the zero-length string state in the cc meson; zero-length string states are allowed in the sc mesons, effectively lowering their potential energy compared to the former. (ii) For fermionic holes, indistinguishability has been shown to introduce additional frustration in the kinetic energy which increases the energy of cc mesons \cite{Trugman1988}. (iii) The spinon kinetic energy of order $\mathcal{O}(J)$ lowers the energy of sc mesons, favoring unpaired states. (iv) Quantum fluctuations tend to lower the effective string tension $\sigma_0$, leading to smaller binding energies. (v) At elevated temperature entropy favors unbound sc mesons. How all these effects add up will be subject of future studies. 

%%%%%%%%%%%%%%%%%%%%%%%%%%%%%%%%%%%%%%
\textbf{Acknowledgements.--}
We thank M. Greiner, J. Koepsell, G. Salomon, C. Gross, S. Hirthe, D. Bourgund, L. Hahn, Z. Zhu, L. Pollet, and U. Schollwöck for fruitful discussions. This research was funded by the Deutsche Forschungsgemeinschaft (DFG, German Research Foundation) under Germany's Excellence Strategy -- EXC-2111 -- 390814868, by the NSF through a grant for the Institute for Theoretical Atomic, Molecular, and Optical Physics at Harvard University and the Smithsonian Astrophysical Observatory, the Harvard-MIT CUA, the ARO grant number W911NF-20-1-0163, and the National Science Foundation through grant No. OAC-1934714.

%%%%%%%%%%%%%%%%%%%%%%%%%%%%%%%%%%%%%%%%%%%%%%%%%%%%%
\section*{Methods}
%%%%%%%%%%%%%%%%%%%%%%%%%%%%%%%%%%%%%%%%%%%%%%%%%%%%%

%%%%%%%%%%%%%%%%%%%%%%%%%%%%%%%%%%%%%%
\textbf{String-based chargon pairing.--}
%%%%%%%%%%%%%%%%%%
Here we introduce the general string-based pairing mechanism for an idealized parton model of AFMIs in $d$ dimensions (the dimer parton model). We consider two flavors of chargons, $\h_{\vec{x},\mu}$ with $\mu=\pm$, which will allow us to work with distinguishable chargons;  $\vec{x} \in \mathbb{Z}^d$ denotes a lattice vector. The two different flavors $\mu=\pm$ could, for example, be internal degrees of freedom or different layers in a bilayer system, see Fig.~\ref{fig1}. 
In the dimer parton model we can treat arbitrary lattice geometries in principle, but we assume a homogeneous coordination number $z$ for all sites; e.g. $z=4$ for the bilayer system shown in Fig.~\ref{fig1} b). Additionally, we introduce $2 \times 2$ flavors of spinons, $\f_{\vec{x},\mu,\sigma}$, where $\sigma=\pm$ denotes the spin index. 

Furthermore, we consider rigid strings $\Sigma$ on the underlying lattice, which connect the partons and fluctuate only through the motion of the latter. In our idealized scenario, we assume that such geometric strings keep a full memory of the parton trajectories up to self-retracing paths, but including loops. These strings can also be viewed as a manifestation of the gauge fluctuations expected from any parton representation of an AFMI.

We work in a regime of low parton density and consider only isolated pairs of two distinguishable partons, $n=1,2$. Hence, their quantum statistics plays no role in the following. The label $n$ summarizes the set of properties flavor $\mu$, parton type (spinon or chargon), and spin $\sigma$ in the case of a spinon. In order to obtain distinguishable partons, at least one of these properties has to be different between $n=1$ and $n=2$. 
We can write the orthonormal basis (ONB) states of the dimer parton model as $\ket{\vec{x}_1,\vec{x}_2,\Sigma}$, where $\Sigma$ is a string connecting $\vec{x}_2$ to $\vec{x}_1$, i.e.
\begin{equation}
 	\bra{\vec{y}_1,\vec{y}_2,\Sigma'} \vec{x}_1,\vec{x}_2,\Sigma \rangle = \delta_{\vec{y}_1,\vec{x}_1} \delta_{\vec{y}_2,\vec{x}_2} \delta_{\Sigma',\Sigma}.
\end{equation}
The effective Hamiltonian of the two partons written in first quantization reads
\begin{multline}
 	\H_{\rm 2p} = - \sum_{n=1}^2 \frac{1}{2 m_n} \sum_{\xy,  \Sigma} \Bigl( \ket{\vec{y}_n,\Sigma'_{\vec{x}_n,\vec{y}_n,\Sigma}} \bra{\vec{x}_n,\Sigma} + \hc \Bigr) \\
 	+ \sum_\Sigma \ket{\Sigma} \bra{\Sigma} ~ V_\Sigma.
 \label{eqH2p}
\end{multline}
The first line describes nearest-neighbor (NN) hopping of the partons, with amplitudes $1/2m_n$, and simultaneous adaption of the string state from $\Sigma$ to $\Sigma'_{\vec{x}_n,\vec{y}_n,\Sigma}$: hopping along the existing string leads to a retraction of the string along $\vec{y}_n-\vec{x}_n$, otherwise the string is extended by the same element. Extensions to next-nearest-neighbor (NNN) hopping can also be considered. The second line in Eq.~\eqref{eqH2p} describes a general string potential $V_\Sigma$ which we assume to be independent of the parton flavors. 

In the following we consider a linear string potential,
\begin{equation}
 	V_\Sigma = \sigma_0 ~ \ell_\Sigma,
\end{equation}
where $\ell_\Sigma$ is the length of string $\Sigma$ and $\sigma_0$ is the linear string tension. In a weakly doped AFMI we may assume that $\sigma_0 \simeq J$ is proportional to the superexchange coupling $J$ \cite{Beran1996,Bohrdt2021arXiv}. Furthermore, we assume that the parton masses strongly depend on flavor in the following way,
\begin{equation}
 	(2t)^{-1} = ~ m_h ~ \ll ~ m_f ~ \simeq J^{-1}.
\end{equation}
I.e. the chargon mass $m_h$ is significantly lighter than the spinon mass $m_f$, since $t \gg t^2/U \simeq J$ in AFMIs.

Our goal in the following is to demonstrate that two distinguishable chargons, i.e. with different flavors $\mu$, form a strongly fluctuating pair with a large binding energy when $t \gg J$. Since we consider rigid strings with a linear string tension, the partons are always confined in our model. Hence the binding energy for two charges is obtained by comparing mesonic states constituted by a spinon-chargon (sc) and a chargon-chargon (cc) pair respectively,
\begin{equation}
 	E_{\rm B} =  2 E_{\rm sc}-E_{\rm cc}.
	\label{eqEBdef}
\end{equation} 

First, we consider sc mesons. Because $1/m_f \simeq J$, we can treat the spinon motion perturbatively and start from a localized spinon. In this case all charge fluctuations are described by NN tunneling between adjacent string configurations, and Eq.~\eqref{eqH2p} becomes a single-particle hopping problem on the Bethe lattice, $\H_{\rm 2p}|_{\rm sc} = - t \sum_{\langle \Sigma', \Sigma \rangle} \l \ket{\Sigma'} \bra{\Sigma} + \hc  \r + \sigma_0 \sum_\Sigma \ell_\Sigma \ket{\Sigma} \bra{\Sigma} + \mathcal{O}(J)$, see Fig.~\ref{fig1} a). Its spectrum is composed of decoupled rotational and vibrational sectors \cite{Grusdt2018tJz}. The ground state has no rotational excitations, and its energy in the limit $t \gg J$ follows the well-known universal form \cite{Bulaevskii1968}
\begin{equation}
 	E_{\rm sc} = - 2 t \sqrt{z-1} + \alpha~  t^{1/3} ~ \sigma_0^{2/3} + \mathcal{O}(J),
	\label{eqEscScaling}
\end{equation}
where $\alpha > 0$ is a non-universal constant proportional to $(z-1)^{1/6}$. Note that contributions from the spinon motion are contained in corrections $\mathcal{O}(J)$.

Next we consider chargon-chargon mesons composed of a pair, $\mu=\pm$, of two distinguishable chargons. Now we must include the dynamics of both partons in Eq.~\eqref{eqH2p}. Since $\H_{\rm 2p}$ is translationally invariant, we can transform to the co-moving frame of the first parton $\vec{x}_1$ by a Lee-Low-Pines transformation \cite{Lee1953}. For a given center-of-mass quasi-momentum $\vec{k}$ from the Brillouin zone (BZ) corresponding to the underlying lattice, one again obtains a hopping problem on the Bethe lattice. While the contributions from the string potential and from chargon $\vec{x}_2$ are identical to the sc case, the motion of the first chargon contributes additional $\vec{k}$-dependent terms which tend to frustrate the motion of the pair for $\vec{k} \neq 0$ \cite{Trugman1988}. 

Around the dispersion minimum at $\vec{k} = 0$ we obtain
\begin{equation}
 	\H_{\rm 2p}|_{cc} = - 2 t \! \! \sum_{\langle \Sigma', \Sigma \rangle} \!\! \l \ket{\Sigma'} \bra{\Sigma} + \hc  \r + \sigma_0 \sum_\Sigma \ell_\Sigma \ket{\Sigma} \bra{\Sigma},
 	\label{eqH2pccMain}
\end{equation}
up to corrections of order $\mathcal{O}(\vec{k}^2)$ \cite{SI}. The only difference to the sc problem is that $t \to 2t$ has been replaced by twice the chargon tunneling. This replacement can be easily understood by noting that the relative motion of the two chargons involves the reduced mass, $1/m_{\rm red} = 2 / m_{h}$, i.e. $t_{\rm red} = 2 t$. Thus we obtain the energy of the chargon-chargon meson by using the same scaling result as in the sc case, Eq.~\eqref{eqEscScaling}, and merely replacing $t \to 2t$. Importantly, this leaves the non-universal constant $\alpha$ unchanged and we obtain
\begin{equation}
	E_{\rm cc} = - 4 t \sqrt{z-1} + \alpha ~ (2 t)^{1/3}~  \sigma_0^{2/3} + \mathcal{O}(J,\vec{k}^2).
	\label{eqEccScaling}
\end{equation}

Now we are in a position to calculate the binding energy, Eq.~\eqref{eqEBdef}, of two holes in the limit $t \gg J$. While the kinetic zero-point energies $\propto - t (z-1)^{1/2}$ cancel, the leading-order string binding energies $\propto t^{1/3} \sigma_0^{2/3}$ yield
\begin{equation}
 	E_{\rm B} = - \alpha ~ \underbrace{(2 - 2^{1/3})}_{=0.740...} ~ t^{1/3} \sigma_0^{2/3} + \mathcal{O}(J). 
 	\label{eqEBstringMeth}
\end{equation}
This is a remarkably strong binding energy, which depends on a combination of $t$ and the linear string tension $\sigma_0 \simeq J$. The appearance of $t$ in this expression highlights the underlying binding mechanism, where two chargons share one string, gaining equal amounts of potential and kinetic energy.

Finally, we estimate the effective mass of the chargon pair on a hypercubic lattice. We make a translationally invariant ansatz for the cc-meson in the co-moving frame of the first chargon, and expand the variational energy up to order $\vec{k}^2$. As shown in the supplementary  \cite{SI}, for $t \gg J$ this yields a center-of-mass dispersion $\vec{k}^2 / 2 M_{\rm cc}$ of the pair, with
\begin{equation}
    M_{\rm cc}^{-1} = 4 t \sqrt{z-1} / z.
    \label{eqMccMeth}
\end{equation}
Despite being tightly bound, the pair is highly mobile -- contrary to common expectations for bipolarons.

%%%%%%%%%%%%%%%%%%%%%%%%%%%%%%%%%%%%%%%%
%\bibliography{/Users/fgrusdt/Documents/Library/dataBase_JabRef2.bib}
%\bibliography{dataBase_JabRef2.bib}

%\bibliography{Library.bib}
%\bibliographystyle{naturemag}%unsrt}

\newpage~
%~\newpage
\onecolumngrid
\appendix 
%%%%%%%%%%%%%%%%%%%%%%%%%%%%%%%%%%%%%%%%%%%%%%%%%%%%%
\section*{Supplementary Information}
%%%%%%%%%%%%%%%%%%%%%%%%%%%%%%%%%%%%%%%%%%%%%%%%%%%%%
In the following appendices, we provide additional details of the parton models discussed in the main text, along with supplementary information about our numerical simulations. 

In Appendix \ref{secApdxTJtoPartons} we explain how the two main parton models we use are related to the microscopic $t-J$ Hamiltonian describing the mixD bilayer systems: We consider (i) the dimer parton model, where we postulate the existence of an ONB basis labeled by the parton and string configurations. This model is obtained from a more realistic set of non-ONB string states by making an ONB approximation and assigning orthogonal states to different dimer coverings of the lattice -- as in the Rokhsar-Kivelson quantum dimer model \cite{Rokhsar1988}. We also consider (ii) the Gram-Schmidt parton model, where the non-ONB string states are properly orthogonalized to obtain a more realistic description of the microscopic $t-J$ model.

In Tab.~\ref{tabPartonModels} we provide an overview how the different parton models are related, and where they are used to obtain different results. 

In Appendix \ref{secApdxDimerPartonModel} we solve the dimer parton model in several interesting regimes. This includes a full solution in $d=1$ and the derivation of the effective mass of the chargon-pair in $d=2$.

In Appendix \ref{secApxGramSchmidt} we define the Gram-Schmidt parton model and show how non-ONB effects in the underlying $t-J$ model introduce spinon dynamics. The resulting Gram-Schmidt parton Hamiltonian we derive can be solved by standard exact diagonalization techniques. 

In Appendix \ref{secApdxDMRG} we provide details of our DMRG calculations (ground state and dynamics) in $d=1$.

%\rowcolors{2}{blue!50}{green!50}
\begin{table*}[h]
\begin{tabular}{@{} *6l @{}}%*5l}    
\toprule
%\emph{name} & \emph{foo} &&&  \\
\rowcolor{gray!30} model name    &    Hilbert space  & Hamiltonian  & Details  & Main results  \\ \midrule
 dimer parton model & dimer-parton-string & Eq.~\eqref{eqH2p} & App. \ref{secApdxDimerPartonModel} & Eq.~\eqref{eqEBstring}, Eq.~\eqref{eqMccMain}, Fig.~\ref{fig:2D} (bottom) \\
 %non-ONB string & $t-J$ & X2 & \ref{secApxSolLadderNonONB} & X4 & bla \\ 
Gram-Schmidt parton model & $t-J$ Hilbert space & Eq.~\eqref{eqHONB} & App. \ref{secApxGramSchmidt} & Fig.~\ref{fig1}c), Fig.~\ref{fig1}, Fig.~\ref{fig:2D} (top), Fig.~\ref{fig:spectra}  \\ 
mixD $t-J$ model & $t-J$ Hilbert space & Eq.~\eqref{eqeqHstring} & App. \ref{secApdxDMRG} & Fig.~\ref{fig1}c), Fig.~\ref{fig1}, Fig.~\ref{fig:2D}, Fig.~\ref{fig:spectra} (colormap)  \\ 
 \bottomrule
 \hline
\end{tabular}
\caption{Overview of the models used in this article.}
\label{tabPartonModels}
\end{table*}

%%%%%%%%%%%%%%%%%%%%%%%%%%%%%%%%%%%%%%%%%%%%%%%%%%%%%
\section{From $t-J$ to the parton models}
\label{secApdxTJtoPartons}
%%%%%%%%%%%%%%%%%%%%%%%%%%%%%%%%%%%%%%%%%%%%%%%%%%%%%
To relate the mixD $t-J$ model with two $d$-dimensional layers ($\mu=\pm$) defined in Eq.~\eqref{eqeqHstring} of the main text to the effective parton model, we first represent the fermions $\c_{\vec{j},\mu,\sigma}$ by slave fermions. I.e. we write
\begin{equation}
    \c_{\vec{j},\mu,\sigma} = \hd_{\vec{j},\mu} \a_{\vec{j},\mu,\sigma}, \quad \sum_\sigma \ad_{\vec{j},\mu,\sigma} \a_{\vec{j},\mu,\sigma} + \hd_{\vec{j},\mu}\h_{\vec{j},\mu} = 1
\end{equation}
with fermionic chargons $\h_{\vec{j},\mu}$ and with Schwinger bosons $\a_{\vec{j},\mu,\sigma}$. Since $\h_{\vec{i},\mu} \hd_{\vec{j},\mu} = - \hd_{\vec{j},\mu} \h_{\vec{i},\mu}$ this leads to the effective hopping Hamiltonian
\begin{equation}
    \H_t = t_\parallel \sum_{\ij} \sum_{\mu, \sigma=\pm} \l \ad_{\vec{i},\mu,\sigma} \a_{\vec{j},\mu,\sigma} \hd_{\vec{j},\mu} \h_{\vec{i},\mu} + \hc \r
    \label{eqHhoppingSlaveFermion}
\end{equation}
with an overall amplitude $+ t_\parallel$. The spin-exchange terms $J_\perp$ and $J_\parallel$ have the usual Schwinger boson representations \cite{Auerbach1998}.

Next we define parton states in the microscopic model. The starting point is the rung-singlet state
\begin{equation}
    \ket{0} = 2^{-V/2} \prod_{\vec{j}} ( \ket{ \!\! \uparrow \downarrow}_{\vec{j}} - \ket{\! \! \downarrow \uparrow}_{\vec{j}} ),
\end{equation}
which is the ground state of the system when $J_\parallel = 0$ and $J_\perp > 0$, at zero doping. Without any strings ($\Sigma=0$), the spinon-chargon and chargon-chargon states are obtained by creating one and two holes, respectively,
\begin{flalign}
    \ket{\vec{j},\mu,\sigma, \Sigma=0}_{\rm sc} = &\hd_{\vec{j},\overline{\mu}} \sqrt{2} \a_{\vec{j},\overline{\mu},\overline{\sigma}} \ket{0}, \\
    \ket{\vec{j}, \Sigma=0}_{\rm cc} = \hd_{\vec{j},+} \hd_{\vec{j},-}& \sum_\sigma \frac{(-1)^\sigma}{\sqrt{2}} \a_{\vec{j},+,\sigma} \a_{\vec{j},-,\overline{\sigma}} \ket{0}.
\end{flalign}
Here $\overline{\mu}=-\mu$ and $\overline{\sigma}=-\sigma$; hence $\vec{j},\mu,\sigma$ denote the quantum numbers of the spinon (sc case).

The string states are then obtained by moving around the chargons along a path $\mathcal{L}_\Sigma$, consisting of a series of links $\ij$ defining the string, and displacing spins along the path. The string operator is
\begin{equation}
    \hat{G}_{\Sigma,\mu} = \prod_{\ij \in \mathcal{L}_\Sigma} \l \hd_{\vec{j},\mu} \h_{\vec{i},\mu} \sum_\sigma \ad_{\vec{i},\mu,\sigma} \a_{\vec{j},\mu,\sigma}\r,
\end{equation}
where the order of the product is important. The corresponding states are
\begin{flalign}
    \ket{\vec{j},\mu,\sigma, \Sigma}_{\rm sc} &= \hat{G}_{\Sigma,\overline{\mu}} \ket{\vec{j},\mu,\sigma, \Sigma=0}_{\rm sc}, \label{eqSCstring}\\
    \ket{\vec{j}_\mu,\Sigma}_{\rm cc} &= \hat{G}_{\Sigma,\overline{\mu}} \ket{\vec{j}, \Sigma=0}_{\rm cc}. \label{eqCCstring}
\end{flalign}
In the chargon-chargon case, the notation $\vec{j}_\mu$ indicates that the chargon in layer $\mu$ is located at site $\vec{j}_\mu$ and the string is introduced by moving the chargon in the opposite layer $\overline{\mu}$. The geometric strings, obtained by displacing spins, correspond to tilted singlets between the two layers $\mu=\pm$, as depicted in Fig.~\ref{fig1}a).

In a ladder, $d=1$, the chargon-chargon states form an orthonormal basis (ONB), $~_{\rm cc}\langle \vec{j}_\mu',\Sigma' \ket{\vec{j}_\mu,\Sigma}_{\rm cc} = \delta_{\Sigma',\Sigma} \delta_{\vec{j}_\mu,\vec{j}_\mu'}$. In a mixD bilayer with $d \geq 2$, two different chargon-chargon states can have identical chargon positions but different strings $\Sigma' \neq \Sigma$. In this case, the basis states are not orthognoal in general, since different singlet coverings have non-zero overlaps in general \cite{Auerbach1998}. Nevertheless, the ONB approximation $\!\!~_{\rm cc}\langle \vec{j}_\mu',\Sigma' \ket{\vec{j}_\mu,\Sigma}_{\rm cc} \approx \delta_{\Sigma',\Sigma} \delta_{\vec{j}_\mu,\vec{j}_\mu'}$, as assumed in the dimer parton model introduced in the main text, is reasonable; for example, two strings $\Sigma, \Sigma'$ whose difference defines a loop around a single plaquette in a square lattice lead to an overlap squared $|\langle \Sigma | \Sigma' \rangle|^2 = 1/16$. If the difference of the two strings defines larger loops, the overlap decays exponentially with the loop size \cite{Auerbach1998}. 

Another limitation of the ONB approximation is provided by loops; in that case, overlaps of distinct string states $\Sigma$ and $\Sigma'$ can even become unity if $\Sigma'$ is obtained from $\Sigma$ by adding loops. For example, in the $d=2$ square lattice starting from $\Sigma$ and performing three complete loops around a single plaquette to obtain $\Sigma'$ restores the original state. I.e. although $\Sigma \neq \Sigma'$, we get $\ket{\vec{j}_\mu,\Sigma'}_{\rm cc} = \ket{\vec{j}_\mu,\Sigma}_{\rm cc}$. However, the fraction of such loop states among all string states with length $\ell=12$ is extremely small: $N_{\rm loop}(\ell=12)/N_{\rm strg}(\ell=12) = 1.1 \times 10^{-5}$. Similarly, other types of loop states have a small relative share of all string states, depending on the geometry and coordination number $z$ of the underlying lattice. This small relative dimension in the Hilbertspace justifies to treat all string states as approximately ONB; effects of loops beyond ONB can be treated systematically in a tight-binding description \cite{Grusdt2018tJz}.

In the sc case, non-ONB effects show up already for a mixD ladder ($d=1$); in this case, moving the spinon by one site, by applying a spin-exchange operator along the leg of the ladder, yields another state which has a non-zero overlap with the initial state. When the spinon hops by one site, the overlap squared is given by
\begin{equation}
    |_{\rm sc}\langle \vec{j}+\vec{e}_x, \mu, \sigma,\Sigma  \ket{\vec{j}, \mu, \sigma,\Sigma }_{\rm sc}|^2 = 1/4.
\end{equation}
When $J_\parallel = 0$, this non-vanishing overlap of neighboring spinon states is solely responsible for the resulting spinon dispersion. Additional $J_\parallel$ couplings introduce further spinon dynamics. A systematic treatment of such non-ONB effects will be provided below, see Sec.~\ref{secApxGramSchmidt}. As in the chargon-chargon case, loops lead to further corrections beyond the ONB prediction, which are expected to be quantitatively small however.

%%%%%%%%%%%%%%%%%%%%%%%%%%%%%%%%%%%%%%%%%%%%%%%%%%%%%
\section{Analytic insights from the dimer parton model}
\label{secApdxDimerPartonModel}
%%%%%%%%%%%%%%%%%%%%%%%%%%%%%%%%%%%%%%%%%%%%%%%%%%%%%
Here we consider ONB states of partons connected by strings (dimer parton Hilbert space) and solve the effective parton Hamiltonian, Eq.~\eqref{eqH2p}.

%%%%%%%%%%%%%%%%%%%%%%%%%%
\subsection{The dimer parton model in $d=1$}
\label{secApxSolLadderONB}
%%%%%%%%%%%%%%%%%%%%%%%%%%
Here we consider the case $J_\parallel=0$ and assume that the geometric string states introduced above, see Eqs.~\eqref{eqSCstring}, \eqref{eqCCstring}, form an ONB. A refined dimer parton Hamiltonian $\H_{\rm rdp}$ can be obtained by calculating matrix elements of the underlying $t-J$ Hamiltonian for the string states. Except for a weak attraction between two chargons at the same site but on opposite layers, this yields the general parton Hamiltonian discussed in the first part of the main text, see Eq.~\eqref{eqH2p}.

\subsubsection{Chargon-chargon case}
Now we show how the (refined) dimer parton model can be solved exactly for $t \gg J$ in $d=1$ dimension. In the following we denote the ONB string basis of the dimer parton Hilbert space by $\ket{x_1,x_2,\Sigma}$ for a pair $\mu=+,-$ of distinguishable chargons at sites $x_{1,2}$, connected by a geometric string. We can drop the index $x_2$ since $x_2 = x_1 + \Sigma$ is determined by the coordinate $x_1$ of the first chargon and the configuration $\Sigma \in \mathbb{Z}$ of the string. The two-chargon Hamiltonian reads
\begin{multline}
 	\H_{\rm rdp} = t_{\parallel} \sum_{\langle \Sigma', \Sigma \rangle} \sum_{x_1} \Bigl( \ket{x_1,\Sigma'}\bra{x_1,\Sigma} +\hc \Bigr)
	+ t_\parallel \sum_{x_1} \sum_\Sigma \Bigl( \ket{x_1 + 1, \Sigma -1} \bra{x_1, \Sigma} + \hc \Bigr) + \\
	+ \frac{J_\perp}{4} \sum_{x_1} \sum_\Sigma \ket{x_1,\Sigma}\bra{x_1,\Sigma} \Bigl( 5 + 3 |\Sigma| -  \delta_{\Sigma,0} \Bigr).
	\label{eqHeff1Dcc}
\end{multline}
The first two terms describe hopping of the two chargons, respectively, and the last term corresponds to the linear string tension. Note that the overall positive sign of the tunneling terms, $+t_\parallel$, is obtained from the string states as described in Sec.~\ref{secApdxTJtoPartons}. By performing a gauge transformation and giving an extra minus sign to odd-length string states, the corresponding Hamiltonian with negative hoppings, $-t_\parallel$, as chosen in the parton model of Eq.~\eqref{eqH2p}, can be obtained.

To solve Eq.~\eqref{eqHeff1Dcc}, we perform a LLP transformation \cite{Lee1953} to the co-moving frame with $x_1$,
\begin{equation}
 	\hat{U}_{\rm LLP} = \exp[- i \hat{x}_1 \hat{p}_2],
\end{equation}
where $\hat{p}_2$ is the momentum of the second chargon. The resulting Hamiltonian 
\begin{equation}
 	\hat{U}_{\rm LLP}^\dagger \H_{\rm rdp} \hat{U}_{\rm LLP} = \sum_k \ket{k}_1\bra{k} ~ \otimes \tilde{\mathcal{H}}_{\rm rdp}(k)
\end{equation}
is block-diagonal in the eigenbasis of $\hat{p}_1$, where each block can be labeled by the total conserved system momentum $k \in [- \pi, \pi)$ in the original basis. The individual blocks take the form
\begin{equation}
 	\tilde{\mathcal{H}}_{\rm rdp}(k) = t_\parallel \sum_\Sigma \Bigl( \ket{\Sigma + 1} \bra{\Sigma} ~ \l  1+ e^{ik} \r+ \hc \Bigr) 
	 + \frac{J_\perp}{4} \sum_\Sigma \ket{\Sigma}\bra{\Sigma}~ \Bigl( 5 + 3 |\Sigma| -  \delta_{\Sigma,0} \Bigr).
	 \label{eqHtildekChCh}
\end{equation}
I.e. in the co-moving frame the effective chargon tunneling is given by
\begin{equation}
 	\tilde{t}(k) = t_\parallel \l  1 + e^{i k} \r
\end{equation}
and depends strongly on $k$. At $k=0$ we obtain $\tilde{t} = 2 t_\parallel$, as anticipated in the main text where we noted that the reduced mass of the chargon pair $1/m_{\rm red} = 2 / m_h$. In contrast, around $k=\pi$ the motion is frustrated by destructive interference of the two chargon tunnelings and we obtain $\tilde{t}(k=\pi) = 0$.

For $|\tilde{t}(k)| \gg J_\perp$ we obtain the following asymptotic scaling of the chargon-chargon spectrum \cite{Bulaevskii1968},
\begin{equation}
 	E_{\rm cc}^{(n)}(k) = E_0 - 2 |\tilde{t}(k)| + \alpha_{\rm cc}^{(n)} |\tilde{t}(k)|^{1/3} J_\perp^{2/3} + \mathcal{O}(J_\perp)
	\label{eqEccScalingAppdx}
\end{equation}
where $n=0,1,2,...$ is the principle quantum number and $E_0$ is the ground state energy of the undoped system; $\alpha_{\rm cc}^{(n)}$ is a non-universal constant increasing with $n$.

%%%%%%%%%%%%%%%%%%%%%%%%%%%%%%%%%%%%%%%%%%%%%%%%%%%%%
\begin{figure}[t!]
\centering
  \includegraphics[width=0.5\linewidth]{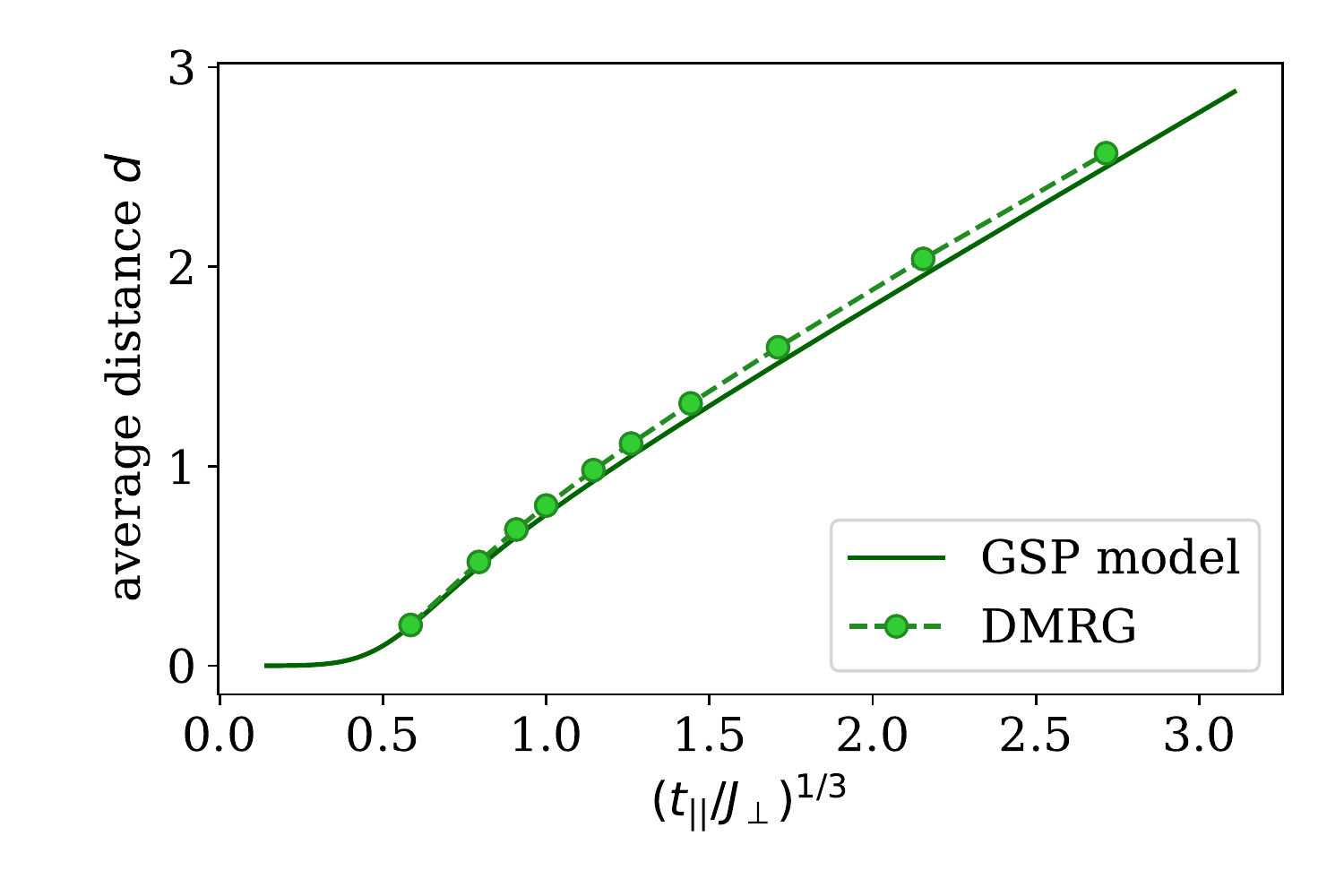}
\caption{\textbf{Average distance between chargons} in the case of two chargons in a mixed dimensional ladder ($d=1$). The average string length obtained from the dimer parton model (solid green line) is compared to the average distance between the two holes obtained from DMRG simulations (green symbols). } 
\label{fig:distances}
\end{figure}
%%%%%%%%%%%%%%%%%%%%%%%%%%%%%%%%%%%%%%%%%%%%%%%%%%%%%

From the solution of the Hamiltonian Eq.~\eqref{eqHeff1Dcc}, the average chargon-chargon string length can be obtained. Similarly, the average distance between the chargons can be obtained from DMRG. As shown in Fig.~\ref{fig:distances}, the resulting distances scale with $(t_{||}/J_\perp)^{1/3}$ for $t_{||}>J_\perp$ as expected. 

\subsubsection{Spinon-chargon case}
A similar calculation can be performed for the sc meson for a given spinon dispersion, $\epsilon_f(k)$, which we will specify further in Sec.~\ref{secApxGramSchmidt}. It is convenient to represent the latter in its Fourier decomposition,
\begin{equation}
 	\epsilon_f(k) = \sum_{r \in \mathbb{Z}_{>0}} J_r e^{i r k} + \hc.
\end{equation}
Transforming to the co-moving frame with the spinon one obtains the dimer parton LLP Hamiltonian
\begin{equation}
 	\tilde{\mathcal{H}}_{\rm dp}(k) =  \sum_{r \in \mathbb{Z}_{>0}} \sum_\Sigma \Bigl( \ket{\Sigma + r} \bra{\Sigma} ~ \l t_\parallel \delta_{r,1} + J_r e^{i r k} \r+ \hc \Bigr) 
	 + \frac{J_\perp}{4} \sum_\Sigma \ket{\Sigma}\bra{\Sigma} ~ \Bigl( 4 + 3 |\Sigma| \Bigr).
	 \label{eqHtildekSpinon}
\end{equation}
Using exact diagonalization techniques, the spectrum of this Hamiltonian can be easily calculated if the maximum string length and the largest $r$ with $J_r \neq 0$ are truncated appropriately.

Neglecting spinon dynamics of order $\mathcal{O}(J_\perp,J_\parallel)$ in Eq.~\eqref{eqHtildekSpinon}, i.e. $J_r \to 0$, we obtain the asymptotic scaling of the sc spectrum for $|t_\parallel| \gg J_\perp \gg  |J_\parallel|$,
\begin{equation}
 	E_{\rm sc}^{(n)}(k) = E_0 - 2 |t_\parallel| + \alpha^{(n)}_{\rm sc} |t_\parallel|^{1/3} J_\perp^{2/3} + \mathcal{O}(J_\perp,J_\parallel)
\end{equation}
where $n=0,1,2,...$ is the principle quantum number and $E_0$ is the ground state energy of the undoped system;

In the absence of the on-site chargon attraction, corresponding to the potential $\propto -J_\perp/4 \cdot \delta_{\Sigma,0}$ in Eq.~\eqref{eqHeff1Dcc}, the non-universal constants in the spinon-chargon and chargon-chargon case are \emph{identical}, $\alpha_{\rm sc}^{(n)} = \alpha_{\rm cc}^{(n)}$; this is the result of the dimer parton model presented in the main text. It leads to strong binding energies $\simeq |t_\parallel |^{1/3} J_\perp^{2/3}$, which justifies neglecting spinon dynamics of $\mathcal{O}(J_\perp,J_\parallel)$ at strong couplings. 

Since the additional interaction $-\delta_{\Sigma,0} J_\perp/4$ considered in the refined dimer parton model is attractive, it follows that $\alpha_{\rm sc}^{(n)} > \alpha_{\rm cc}^{(n)}$ and binding of chargons is favored further by this term.

%%%%%%%%%%%%%%%%%%%%%%%%%%%%%%%%%%%%%%%%%%%%%%%%%%%%%
\subsection{Solution of the dimer parton model in 2D: Two chargons at $\mathbf{k}=0$}
\label{secApdxDimerStringTwoPartonsk0}
%%%%%%%%%%%%%%%%%%%%%%%%%%%%%%%%%%%%%%%%%%%%%%%%%%%%%
Here we solve the dimer parton Hamiltonian from Eq.~\eqref{eqH2p} in the main text with a general string potential depending only on the string length, for two distinguishable chargons with total momentum $\vec{k} \approx 0$. Unlike in the case of a sc pair, where the internal rotational symmetry around the heavy spinon remains intact for all $\vec{k}$ when $t \gg J$, both partons are mobile in the chargon-chargon case. Hence, their relative motion couples to the center-of-mass motion in the lattice, and the internal rotational symmetry of the meson is lost. However, for rotationally-invariant  momenta (RIM) $\vec{k}_{\rm RIM}$ which respect the discrete rotational symmetry of the underlying lattice, the relative motion of the chargons remains rotationally invariant. Formally this can be seen by performing a Lee-Low-Pines (LLP) transformation \cite{Lee1953} into a frame co-moving with one of the partons, see Ref.~\cite{Bohrdt2021arXiv} for a general argument or Sec.~\ref{secApxSolLadderONB} for an explicit calculation in $d=1$.

In the following, we restrict our analysis to the RIM $\vec{k}=0$. This momentum is particularly important, since it corresponds to the lowest-energy chargon-chargon state. To see this, we note that a LLP transformation into the co-moving frame with parton $n=1$ yields tunneling terms of strength \cite{Bohrdt2021arXiv} $\tilde{t}(\vec{k}) = - \nicefrac{1}{2} [ m_1^{-1} + e^{i \vec{k} \cdot \vec{e}_\nu} m_2^{-1}]$, where $\nu$ denotes hopping directions $\vec{e}_\nu$ in the lattice. Moreover these are the only $\vec{k}$-dependent terms in the LLP Hamiltonian. Hence, when $m_1=m_2$ as in the case of two distinguishable but otherwise identical chargons, the kinetic energy is minimized for $\vec{k}=0$ and quenched for $\vec{k}=\vec{\pi}$, i.e. $\tilde{t}(\vec{\pi})=0$. This generic behavior of chargon-chargon states is confirmed in our explicit LLP calculation for $d=1$ in Sec.~\ref{secApxSolLadderONB} above. Note that the sign of $m_1=m_2 = m$ does not matter here, since gauge transformations of the string states in the LLP frame allow changing it from $m \to - m$.

To describe the chargon-chargon ground state around $\vec{k}\approx 0$, we make the following ansatz for the chargon-chargon wavefunction:
\begin{equation}
    \ket{\Psi_{\rm cc}(\vec{k})} = L^{-d/2} \sum_{\vec{x}_+} e^{i \vec{k} \cdot \vec{x}_+}  \sum_{\ell=0}^\infty \psi_{\ell} \sum_{\Sigma: \ell_\Sigma = \ell} \ket{\vec{x}_+,\Sigma}.
    \label{eqHcck}
\end{equation}
Here $d$ is the dimension of the lattice with volume $L^d$; we chose to define the string in the co-moving frame with the chargon $\mu=+$ at site $\vec{x}_+$. The sums in Eq.~\eqref{eqHcck} include all lattice sites $\vec{x}_+$ and all string configurations $\Sigma$ of length $\ell_\Sigma = \ell$, each with amplitude $\psi_\ell$. Note the absence of any rotational excitation of the string \cite{Grusdt2018tJz,Bohrdt2021arXiv}; all string directions have equal weights $\psi_\ell$ depending only on the string length.

Before we proceed, we note that each string length $\ell$ corresponds to $\mathcal{N}_\ell$ orthogonal string configurations. Their number only depends on the coordination number $z$ of the lattice: $\mathcal{N}_0=1$, and for $\ell \geq 1$: 
\begin{equation}
    \mathcal{N}_\ell = z ~ (z-1)^{\ell-1}.
    \label{eqNell}
\end{equation}
In Eq.~\eqref{eqHcck} only the symmetric superposition of all these configurations appears, which properly normalized reads
\begin{equation}
    \ket{\ell} = \mathcal{N}_\ell^{-1/2} \sum_{\Sigma: \ell_\Sigma = \ell} \ket{\Sigma},
    \label{eqStringLengthBasis}
\end{equation}
with $\langle \ell' \ket{\ell} = \delta_{\ell',\ell}$. I.e. defining $\phi_\ell = \mathcal{N}_{\ell}^{1/2} \psi_\ell$, the ansatz from Eq.~\eqref{eqHcck} becomes
\begin{equation}
    \ket{\Psi_{\rm cc}(\vec{k})} = L^{-d/2} \sum_{\vec{x}_+} e^{i \vec{k} \cdot \vec{x}_+} \ket{\vec{x}_+} \sum_{\ell=0}^\infty \phi_{\ell} \ket{\ell}.
    \label{eqPsicckSymm}
\end{equation}
Normalization requires $\sum_{\ell=0}^\infty |\phi_\ell|^2 = 1$.

Now we consider the case $\vec{k}=0$. We first note that $\ket{\Psi_{\rm cc}(\vec{0})}$ is an exact eigenstate of the chargon-chargon Hamiltonian $\H_{\rm 2p}$, Eq.~\eqref{eqH2p} from the main text, with a string potential $V_\Sigma = V(\ell_\Sigma)$ depending only on the string length $\ell_\Sigma$, if $\phi_\ell$ is an eigenstate of the single-particle Hamiltonian
\begin{equation}
  \H_{\rm eff} = - \frac{\sqrt{z}}{m} \Bigl( \ket{\ell=1}\bra{\ell=0} + \hc \Bigr) 
  - \frac{\sqrt{z-1}}{m} \sum_{\ell = 1}^\infty \Bigl( \ket{\ell+1}\bra{\ell} + \hc \Bigr)  
  + \sum_{\ell=0}^\infty V(\ell) \ket{\ell}\bra{\ell}.
  \label{eqHeffcck0}
\end{equation}
Here $m=m_+ = m_-$ is the mass of the chargons; the eigenstates are of the form $\ket{\phi} = \sum_{\ell=0}^\infty \phi_\ell \ket{\ell}$ and $\ket{\ell}$ denote the orthonormal basis of symmetric string-length states. The eigenenergy $E_{\rm cc}$ of this single-particle Hamiltonian gives the correct eigenenergy of $\H_{\rm 2p}$. From here, it is straightforward to obtain the results in Eqs.~\eqref{eqH2pccMain}, \eqref{eqEccScaling} of the main text \cite{Bulaevskii1968}. 

For $\vec{k} \neq 0$, we can make the same ansatz as in Eq.~\eqref{eqPsicckSymm} which includes only equal superpositions of different string configurations with the same length. In this case, the ansatz is no longer an exact eigenstate of $\H_{\rm 2p}$, but we can still use it as a reasonable variational wavefunction around $\vec{k} \approx 0$. Assuming a hypercubic lattice in $d$ dimensions, with coordination number $z=2d$, the corresponding variational energy can be evaluated as
\begin{equation}
    \bra{\Psi_{\rm cc}(\vec{k})} \H_{\rm 2p} \ket{\Psi_{\rm cc}(\vec{k})} = E_{\rm cc}(0) + \frac{\vec{k}^2}{2 M_{\rm cc}} + \mathcal{O}(\vec{k}^4).
    \label{eqEvark}
\end{equation}
Here $E_{\rm cc}(0)$ is the exact result at $\vec{k}=0$ and the effective mass of the pair is given by
\begin{equation}
    M_{\rm cc} = \frac{z}{2 t \sqrt{z-1} ~ \bra{\phi}\H_{\rm kin} \ket{\phi}} ,
    \label{eqMccEstimate}
\end{equation}
where the state $\ket{\phi}$ was defined above and
\begin{equation}
    \H_{\rm kin} = \frac{\sqrt{z}}{\sqrt{z-1}} \Bigl( \ket{\ell=1}\bra{\ell=0} + \hc \Bigr) 
  + \sum_{\ell = 1}^\infty \Bigl( \ket{\ell+1}\bra{\ell} + \hc \Bigr)
\end{equation}
is the kinetic part of the effective Hamiltonian defined in Eq.~\eqref{eqHeffcck0}. The proof of Eq.~\eqref{eqEvark} is a tedious but in principle straightforward calculation. When $t \gg J$, one obtains $\bra{\phi}\H_{\rm kin} \ket{\phi} = 2$, which leads to the estimate for the mass presented in the main text.

We emphasize that the result in Eq.~\eqref{eqMccEstimate} is a variational estimate. For $\vec{k} \neq 0$ the string part of the ansatz wavefunction in Eq.~\eqref{eqHcck}, $\psi_\ell$, remains independent of $\vec{k}$ in our ansatz. By adapting $\psi_\ell \to \psi_\ell(\vec{k})$ improved estimates for the effective mass can be obtained. However, the simpler variational approximation above captures the main features of the effective meson mass $M_{\rm cc}$, in particular its scaling with $t^{-1}$ when $t \gg J$. This can be confirmed by an exact calculation of $M_{\rm cc}$ in the limit $t \gg J$ in $d=1$ dimension, see Sec.~\ref{secApxSolLadderONB}. There a LLP transformation leads to the exact asymptotic result $M_{\rm cc} |_{\rm exact} = 1/t$, whereas our estimate, see Eq.~\eqref{eqMccMain} in the main text, gives $M_{\rm cc} |_{\rm est} = 1/(2t)$, which is off by a factor of $2$.

%%%%%%%%%%%%%%%%%%%%%%%%%%%%%%%%%%%%%%%%%%%%%%%%%%%%%
\section{Gram-Schmidt parton model}
\label{secApxGramSchmidt}
%%%%%%%%%%%%%%%%%%%%%%%%%%%%%%%%%%%%%%%%%%%%%%%%%%%%%

%%%%%%%%%%%%%%%%%%%%%%%%%%
%\subsection{Beyond the string model: spinon dynamics}
%\label{secApxSolLadderNonONB}
%%%%%%%%%%%%%%%%%%%%%%%%%%
Now we go beyond the dimer parton model and describe the effects of non-orthogonal string states in the microscopic mixD bilayer $t-J$ model, Eq.~\eqref{eqeqHstring} in the main text, using the Gram-Schmidt method. This leads us to a microscopic description of spinon dynamics in the system. We show that taking into account the following two corrections to the dimer parton model gives a significant quantitative improvement and leads to excellent agreement with our numerical DMRG calculations in $d=1$:
\begin{itemize}
    \item[1.)] non-orthogonality between different parton and string states (not relevant for the chargon-chargon bound state in the ladder), and
    \item[2.)] fluctuations of the string due to magnetic interactions~$J_\perp$ between the layers and~$J_\parallel$ within one layer (see Eq.~(\ref{eqeqHstring})).
\end{itemize}
Moreover, we will work in the co-moving frame of the chargon to describe the parton dynamics of the spinon-chargon (chargon-chargon) pairs.

%%%%%%%%%%%%%%%%%%%%%%%%%%%%%%%%%%%%%%%%%%%%%%%%%%%%%
\subsection{Gram-Schmidt parton model in the ladder ($d=1$)}
\label{secApxGSPLadder}
%%%%%%%%%%%%%%%%%%%%%%%%%%%%%%%%%%%%%%%%%%%%%%%%%%%%%
First, we note that chargon-chargon string states~$\ket{j_\mu,\Sigma}_{\rm cc}$ in a ladder fulfill the orthogonality condition $_{\rm cc}\bra{j'_{\mu},\Sigma'} j_\mu,\Sigma\rangle_{\rm cc} = \delta_{j,j'}\delta_{\Sigma,\Sigma'}$ and thus Hamiltonian Eq.~(\ref{eqHtildekChCh}) in the co-moving frame does not need to be modified due to non-ONB effects.
Further, magnetic interactions~$J_\parallel$ within the layer only yield a constant energy shift but do not add additional dispersion to neither the chargon motion nor parton dispersion in the approximate string basis.

Next, we consider the spinon-chargon (sc) string states~$\ket{j_{\rm c},\mu,\sigma, \Sigma}_{\rm sc}$; here $j_{\rm c}$ denotes the location of the chargon, and the string $\Sigma \in \mathbb{Z}$ points from the chargon to the spinon. This convention is different from the notation in Eq.~\eqref{eqSCstring}, but it is advantageous since two states with chargons at different positions $j_{\rm c}' \neq j_{\rm c}$ are orthogonal. In general, the overlaps of these sc states are
\begin{equation}
    _{\rm sc}\bkt{j_{\rm c}',\mu',\sigma', \Sigma'}{j_{\rm c},\mu,\sigma, \Sigma}_{\rm sc} = \delta_{j_{\rm c},j_{\rm c}'}\delta_{\mu,\mu'}\delta_{\sigma,\sigma'} g_{\Sigma,\Sigma'}
    \label{eqOverlaps}
\end{equation}
where we defined a metric tensor~$g$ with matrix elements
\begin{equation}
g_{\Sigma,\Sigma'}= 2^{-|\Sigma-\Sigma'|}.
\label{eqMetric}
\end{equation}
To deal with the non-orthonormality, we first project the Hamiltonian onto a subspace of the full Hilbertspace spanned by the string states $\ket{j_{\rm c},\Sigma}$. In particular, we derive the matrix elements of the projected Hamiltonian through a variational ansatz $\ket{\Psi^{(n)}(k)}$ defined below. 

We then perform a basis transformation to an ONB by means of a Gram-Schmidt procedure.
We thus obtain an ONB Hamiltonian, which we call the Gram-Schmidt parton (GSP) model. The GSP model can then be treated by exact diagonalization methods.

We start with a translationally-invariant variational ansatz of strings fluctuating around the chargon,
\begin{equation}
    \ket{\Psi^{(n)}} = L^{-1/2} \sum_{j_{\rm c}} e^{i k j_{\rm c}} \sum_{\Sigma} \psi^{(n)}_{\Sigma} \ket{j_{\rm c}, \Sigma},
    \label{eqVarAnsatzLadder}
\end{equation}
which is the most general ansatz respecting the total momentum conservation.
Furthermore, without loss of generality, we have chosen to work in the sector~$\ket{j_{\rm c}, \Sigma}\equiv\ket{j_{\rm c},\mu,\sigma, \Sigma}_{\rm sc} = \ket{j_{\rm c},\mu=+1,\sigma=\uparrow, \Sigma}$.
This is justified because we consider no tunneling between the layers, $t_\perp=0$, and the Hamiltonian conserves total~$\hat{S}^z = \sum_{j,\mu}\hat{S}^z_{j,\mu}$.
The variational kinetic energy as per Hamiltonian~Eq.~(\ref{eqHhoppingSlaveFermion}) yields
\begin{align}
    \langle \H_t (k) \rangle =  \frac{t_\parallel}{L}&\sum_{j_{\rm c},j_{\rm c}'}e^{i k ( j_{\rm c} - j_{\rm c}')}\sum_{\Sigma,\Sigma'} \bar{\psi}^{(n)}_{\Sigma'}\psi^{(n)}_{\Sigma} \bra{j_{\rm c}',\Sigma'}\bigg(\ket{j_{\rm c}+1,\Sigma-1}+\ket{j_{\rm c}-1,\Sigma+1}\bigg).
\end{align}
By shifting the summation indices $j_{\rm c}$ and $j_{\rm c}'$ and using Eqs.~\eqref{eqOverlaps},~\eqref{eqMetric} we find
\begin{align}
\begin{split}
    \langle \H_t (k) \rangle =  t_\parallel \sum_{\Sigma,\Sigma'} \bar{\psi}^{(n)}_{\Sigma'}\psi^{(n)}_{\Sigma}
    %\underbrace{
    (g_{\Sigma',\Sigma-1}e^{-ik} + g_{\Sigma',\Sigma+1}e^{ik})
    %}_{t^{-1}_{\parallel}\sum_{\Sigma''}g_{\Sigma',\Sigma''}h^t_{\Sigma'',\Sigma}(k)},
\end{split}
\end{align}
and  define matrix elements $h^t_{\Sigma'',\Sigma}(k)$ in the non-ONB basis as
\begin{equation}
    \sum_{\Sigma''}g_{\Sigma',\Sigma''}h^t_{\Sigma'',\Sigma}(k) = t_{\parallel} (g_{\Sigma',\Sigma-1}e^{-ik} + g_{\Sigma',\Sigma+1}e^{ik}).
\end{equation}
The metric tensor~$\hat{g}$ is involved when evaluating the scalar product~$\bra{\Psi^{(n)}} \H_t \ket{\Psi^{(n)}}$ and thus appears in the expression for the variational energy. Later, we want to extract the matrix $h^t_{\Sigma',\Sigma}$ in order to derive an effective Hamiltonian.
In particular, the kinetic part of the Hamiltonian yields all-to-all interactions in the non-ONB with exponentially decaying, complex hopping amplitudes, see Eq.~\eqref{eqOverlaps}.

Now, we introduce the basis transformation~$\hat{\mathcal{G}}$ that maps from the non-ONB into a Gram-Schmidt orthonormalized basis $\ket{j_{\rm c},\tilde{\Sigma}}$ with
\begin{align}
    \ket{j_{\rm c},\Sigma} &= \sum_{\tilde{\Sigma}}\mathcal{G}_{\tilde{\Sigma},\Sigma}\ket{j_{\rm c},\tilde{\Sigma}}\\
    \ket{j_{\rm c},\Sigma=0} &= \ket{j_{\rm c},\tilde{\Sigma}=0}.
    \label{eqGramSchmidt}
\end{align}
Since the variational kinetic energy $\langle \H_t (k) \rangle$ must be invariant under basis transformations, we can deduce an effective hopping Hamiltonian~$\H^{\rm GSP}_t=\hat{\mathcal{G}}\hat{h}^t\hat{\mathcal{G}}^{-1}$ in the Gram-Schmidt ONB which takes the particular form
\begin{align}
    (\H^{\rm GSP}_t)_{\tilde{\Sigma}',\tilde{\Sigma}} = &t_\parallel \sum_{\Sigma,\Sigma',\Sigma''}\mathcal{G}_{\tilde{\Sigma}',\Sigma''} (g^{-1})_{\Sigma'',\Sigma'}\big(g_{\Sigma',\Sigma-1}e^{-ik} + g_{\Sigma',\Sigma+1}e^{ik}\big) (\mathcal{G}^{-1})_{\Sigma,\tilde{\Sigma}}
\end{align}
The advantage of having expressions for the matrix elements in an ONB is that standard exact diagonalization (ED) techniques and algorithms can be applied; the resulting spectrum can be directly compared to our DMRG calculations.

Similar as for the kinetic term, we can derive the magnetic interactions of the Gram-Schmidt parton model, i.e.~$\H^{\rm GSP}_{J_\perp}$ and~$\H^{\rm GSP}_{J_\parallel}$.
Here, it is convenient to re-write the nearest neighbour spin-spin interaction by~$\hat{\vec{S}}_{i,\mu}\cdot\hat{\vec{S}}_{j,\mu'}=1/2\hat{P}_{(i,\mu),(j,\mu')}-1/4$, where~$\hat{P}_{(i,\mu),(j,\mu')}$ permutes the spins on sites $(i,\mu)$ and $(j,\mu')$.
Applying the spin-spin interaction on the variational wavefunction Eq.~(\ref{eqVarAnsatzLadder}), i.e.~$\ket{\varphi^{(n)}}= \H_{J}\ket{\Psi^{(n)}}$, maps onto states that are not included in the subspace spanned by $\{\ket{j_{\rm c},\Sigma} \}_{j_{\rm j}\in \mathbb{Z},\Sigma \in \mathbb{Z}}$. Hence, the evaluation of the variational energy requires to calculate overlaps that arise from the projection~$\langle \Psi^{(n)} \ket{\varphi^{(n)}}$ back onto the subspace of interest.

Carefully calculating the variational energies by taking the appropriate overlaps into account yields the following matrix elements in the non-ONB basis:
\begin{align}
\begin{split}
    h^{J_\perp}_{\Sigma',\Sigma} &= -J_\perp\left(L-1-\frac{|\Sigma|}{2}\right) + \frac{J_\perp}{2}\sum_{0 < \xi \le |\Sigma|} f^\xi_{\Sigma',\Sigma} \\
    f^\xi_{\Sigma',\Sigma} &= \begin{cases} \frac{1}{2}, & \text{for}~~(|\Sigma'|\ge \xi) \wedge (\text{sgn}[\Sigma']=\text{sgn}[\Sigma]) \\ -1, & \text{else}\end{cases} 
\end{split}
\end{align}
and
\begin{align}
\begin{split}
    h^{J_\parallel}_{\Sigma',\Sigma} &= \frac{J_\parallel}{2t_\parallel}h^{t}_{\Sigma',\Sigma}(k=0) - J_\parallel(L - 3 + \delta_{\Sigma,0})\\
    & + J_\parallel\sum_{0< \xi < \ell_\Sigma-1}F^{\xi}_{\Sigma',\Sigma} \\
    & + J_\parallel\sum_{\substack{\chi\notin \{0,..,\Sigma+1 \} \\ \text{if}~\Sigma>0} }F^{\chi}_{\Sigma',\Sigma} + J_\parallel\sum_{\substack{\chi\notin \{0,..,\Sigma-1 \} \\ \text{if}~\Sigma<0} } F^{\chi}_{\Sigma',\Sigma} \\
    F^{\xi}_{\Sigma',\Sigma} &= \begin{cases} \frac{1}{2}, & \text{for}~~(|\Sigma'|\ge \xi) \wedge (\text{sgn}[\Sigma']=\text{sgn}[\Sigma]) \\ 2, & \text{else.}\end{cases} \\
    F^{\chi}_{\Sigma',\Sigma} &= \begin{cases} 2, & \text{for}~~(|\Sigma'| > \xi) \wedge (\text{sgn}[\Sigma']=\text{sgn}[\Sigma]) \\ \frac{1}{2}, & \text{else.}\end{cases} 
\end{split}
\end{align}

Note that after transforming into the ONB basis, the operators acquire a hermitian form.
Moreover, the complete effective ONB Hamiltonian
\begin{align}
    \H^{\rm GSP} = \hat{\mathcal{G}}(h^t + h^{J_\perp} + h^{J_\parallel})\hat{\mathcal{G}}^{-1}
    \label{eqHONB}
\end{align}
can be exactly diagonalized, where the string basis Hilbertspace is truncated at some maximum string length~$\ell_\Sigma^{\rm max}$.
Since in the $d=1$ ladder the Hilbertspace dimension only grows linearly with system size, this is not a limiting factor in the calculations and for~$\ell_\Sigma^{\rm max}=25$ the ground state energy converges already sufficiently for the sets of parameters we used in the main text.

\begin{figure}[t!]
\centering
\subfigure{
\centering
  \includegraphics[width=0.47\textwidth]{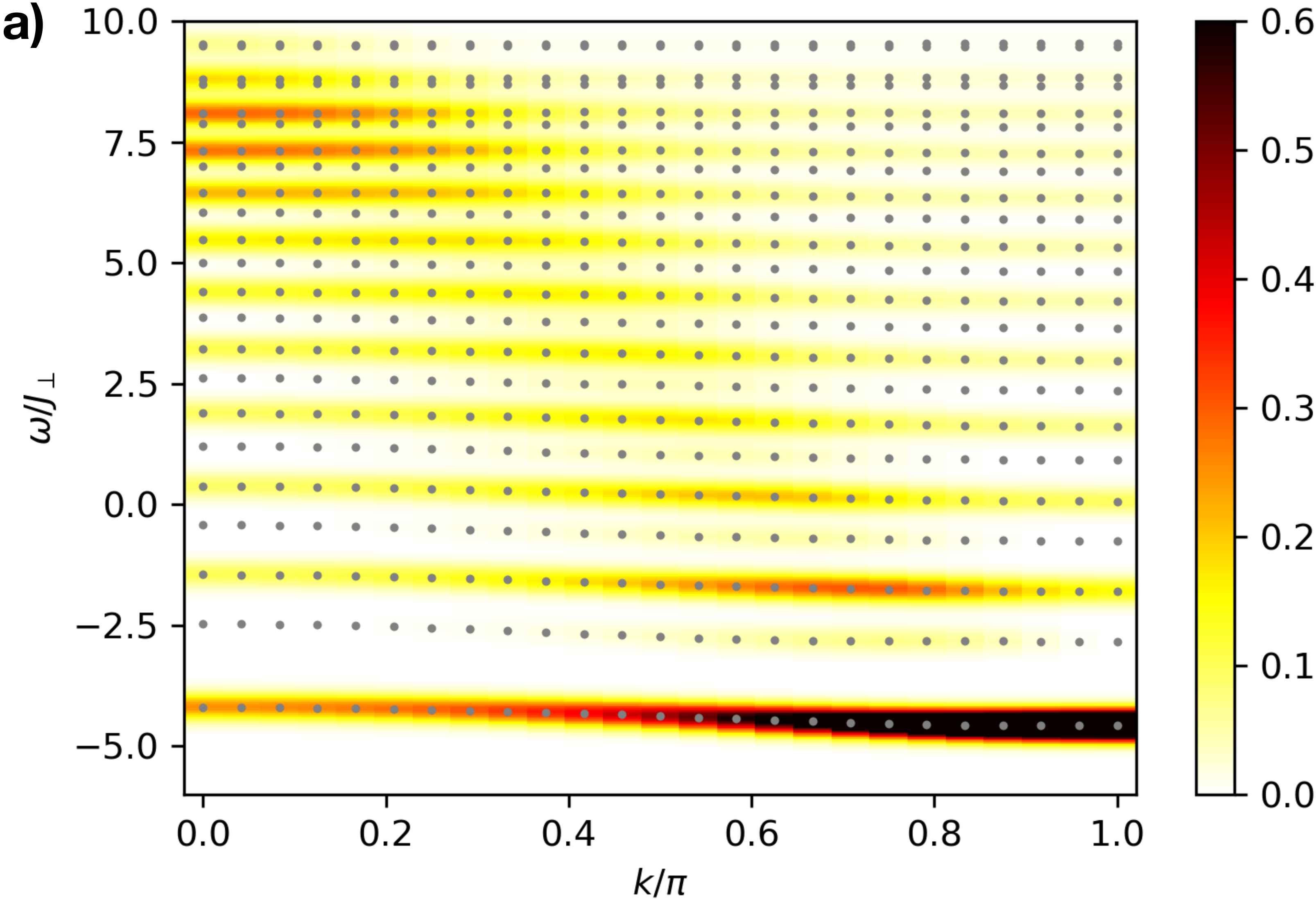}
}
\hfill
\subfigure{
\centering
  \includegraphics[width=0.47\textwidth]{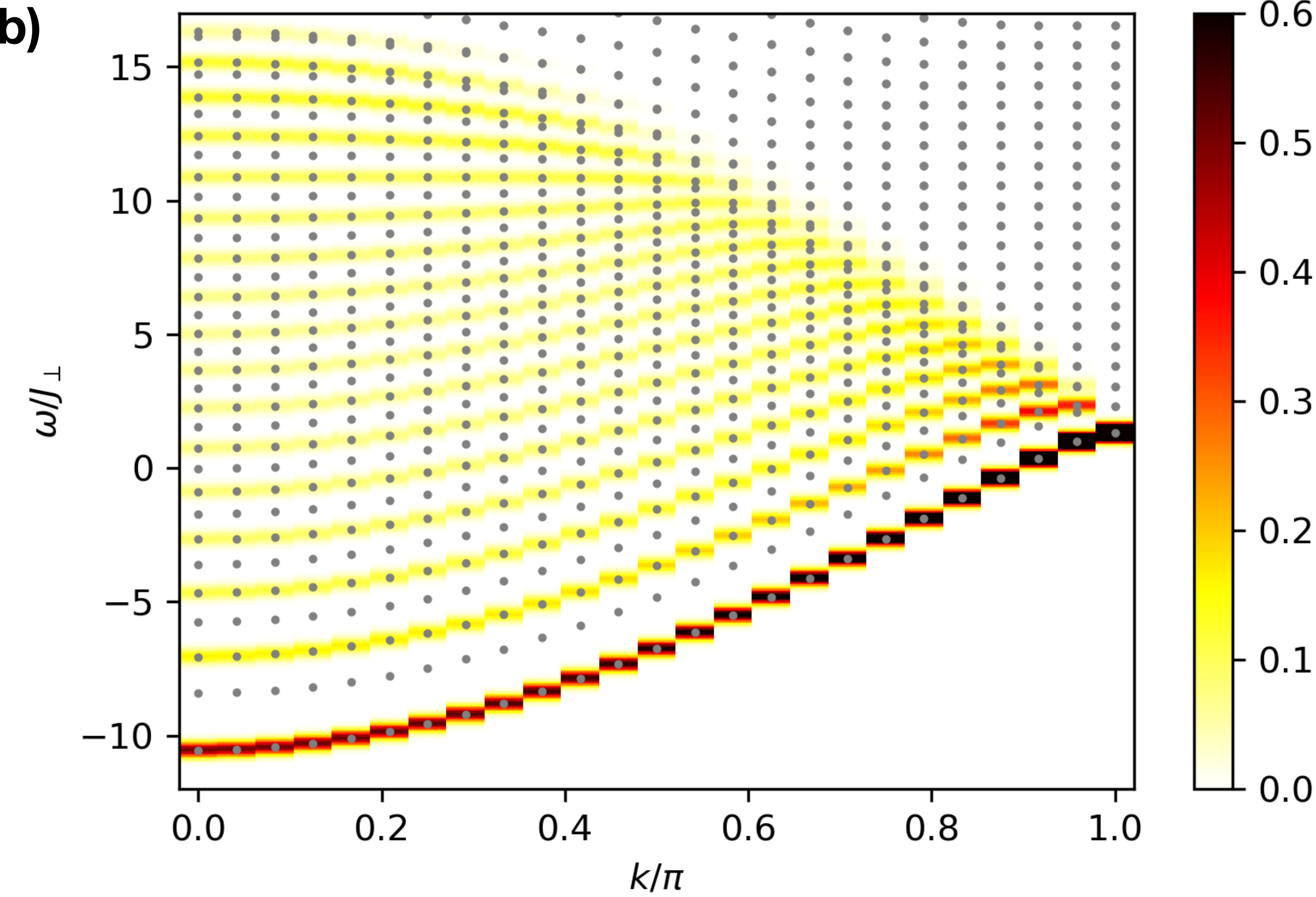}
}
\caption{\textbf{One- and two-hole spectra} in a mixD ladder obtained from ED after performing a Gram-Schmidt orthogonalization procedure and taking into account the spinon dynamics from $J_\parallel$ magnetic interactions for $t_\parallel/J_\perp = 3.4$ and $J_\parallel/J_\perp=0.31$. Parameters are identical to those in Fig.~\ref{fig:spectra}a), b) in the main text.} 
\label{fig:EDspectralfunction}
\end{figure}

To meaningfully compare the ED results with the spectral function obtained by DMRG calculations, we need to subtract the zero-hole energy~$E_{0h}$ given by
\begin{align}
    E_{0h} = -J_\perp L - \frac{J_\parallel}{2} (L-1).
\end{align}
Moreover, the ED calculations also give direct access to the eigenstates of Hamiltonian~(\ref{eqHONB}) from which we can calculate the spectral function (see~Eq.~(\ref{eqSpectralFunction})).
To account for the finite time-evolution used in our DMRG simulations, and thus broadening of the spectral peaks, we multiply the spectral function obtained from ED with a Gaussian of width~$W/\omega = 0.2$. Our results are shown in Fig.~\ref{fig:EDspectralfunction}.

From the energy spectra for spinon-chargon and chargon-chargon mesons we can extract the binding energy as discussed in the main text and as shown in Fig.~\ref{fig1} of the main text.

%%%%%%%%%%%%%%%%%%%%%%%%%%%%%%%%%%%%%%%%%%%%%%%%%%%%%
\begin{comment}
\begin{figure}[t!]
\centering
  \includegraphics[width=0.99\linewidth]{ED_spectralfunction.pdf}
\caption{\textbf{One- and two-hole spectra} in a mixD ladder obtained from ED after performing a Gram-Schmidt orthogonalization procedure and taking into account the spinon dynamics from $J_\parallel$ magnetic interactions for $t_\parallel/J_\perp = 3.4$ and $J_\parallel/J_\perp=0.31$. Parameters are identical to those in Fig.~\ref{fig:spectra}a), b) in the main text.} 
\label{fig:EDspectralfunction}
\end{figure}
\end{comment}
%%%%%%%%%%%%%%%%%%%%%%%%%%%%%%%%%%%%%%%%%%%%%%%%%%%%%

%%%%%%%%%%%%%%%%%%%%%%%%%%%%%%%%%%%%%%%%%%%%%%%%%%%%%
%%%%%%%%%%%%%%%%%%%%%%%%%%%%%%%%%%%%%%%%%%%%%%%%%%%%%

%%%%%%%%%%%%%%%%%%%%%%%%%%%%%%%%%%%%%%%%%%%%%%%%%%%%%
\subsection{Non-ONB effects in the mixD bilayer model ($d=2$)}
%%%%%%%%%%%%%%%%%%%%%%%%%%%%%%%%%%%%%%%%%%%%%%%%%%%%%
In $d=2$ the number of string basis states~$\ket{\vec{j}_{\rm c},\mu,\sigma, \Sigma}_{\rm sc}$ and $\ket{\vec{j}_{\mu}, \Sigma}_{\rm cc}$, respectively, grows exponentially with the length of the string~$\ell_\Sigma$ and thus becomes intractable to be solved using exact diagonalization.
In the scope of this paper, we are only interested in the binding energy and hence in the ground-state energy of the spinon-chargon and chargon-chargon case, which was argued to be at RIM $\vec{k}=0,\vec{\pi}$ in Sec.~(\ref{secApdxDimerStringTwoPartonsk0}).
Hence, working in the co-moving frame of the chargon, we can make an ansatz in the string length basis~$\ket{\ell}$ Eq.~(\ref{eqStringLengthBasis}), which assumes a symmetric superposition of all strings with equal given length $\ell$. The ansatz cannot capture any rotational excitation but assumes a state with no angular momentum, i.e.\ s-wave.

Ignoring loop effects, the chargon-chargon case was discussed in Sec.~(\ref{secApdxDimerStringTwoPartonsk0}). Because two string length states~$\ket{\ell}$ and $\ket{\ell'}$ have different chargon positions (ignoring loops), the basis forms an ONB and the chargon-chargon ground-state energy can be determined by diagonalizing Eq.~(\ref{eqHeffcck0}).

The more involved problem is the sc case, for which we want to find the ground-state energy for~$|t_\parallel|,|J_\perp|>0$ and~$J_\parallel=0$. As in Sec.~(\ref{secApdxDimerStringTwoPartonsk0}), we want to work in the string length basis~$\ket{\ell}_{\rm sc}$, however different string states are now not mutually orthogonal but
\begin{align}
    _{\rm sc}\langle \ell' \ket{\ell}_{\rm sc} &= \mathcal{N}_{\ell'}^{-1/2} \mathcal{N}_{\ell}^{-1/2} \sum_{\Sigma': \ell_{\Sigma'} = \ell'} \sum_{\Sigma: \ell_{\Sigma} = \ell} ~_{\rm sc}\langle \vec{j}_{\rm c}, \Sigma' \ket{\vec{j}_{\rm c}, \Sigma}_{\rm sc} = g^{d=2}_{\ell',\ell}.
\end{align}
Here, we have chosen to work in the sector~$\ket{\vec{j}_{\rm c}, \Sigma}_{\rm sc} = \ket{\vec{j}_{\rm c}, \mu=+1, \sigma=\uparrow, \Sigma}_{\rm sc}$. Further we have calculated the metric tensor

\begin{align}
    g^{d=2}_{\ell',\ell} = \begin{cases} \big(\mathcal{N}_{\ell}^{1/2}2^{-|\ell|}\big)\big( \mathcal{N}_{\ell'}^{1/2}2^{-|\ell'|} \big), & \text{for}~~\ell=0 \vee \ell'=0 \\
    \mathcal{N}_{\ell'}^{-1/2}\mathcal{N}_{\ell}^{1/2}\bigg[ 2^{-|\ell-\ell'|} + \sum\limits_{\lambda \geq 1 }^{\ell'-1} (z-2)(z-1)^{\lambda-1}2^{-|\ell-\ell'|-2\lambda} + \mathcal{N}_{\ell'}\dfrac{z-1}{z}2^{-|\ell+\ell'|} \bigg], & \text{for}~~\ell \geq \ell' \\
    \mathcal{N}_{\ell'}^{-1/2}\mathcal{N}_{\ell}^{1/2}\bigg[ (z-1)^{\ell'-\ell}2^{-|\ell-\ell'|} + \sum\limits_{\lambda \geq 1}^{\ell-1} (z-2)(z-1)^{\lambda-1+\ell'-\ell}2^{-|\ell-\ell'|-2\lambda}+\mathcal{N}_{\ell'}\dfrac{z-1}{z}2^{-|\ell+\ell'|} \bigg], &\text{else}.
    \end{cases}
\end{align}

The combinatorial exercise is convenient to be examined on a Bethe lattice as shown in Fig.~\ref{fig1}; $z$ is the coordination number of the parent lattice and~$\mathcal{N}_{\ell}$ is defined in Eq.~(\ref{eqNell}).
Note that the states~$\ket{\ell}_{\rm sc}$ are not normalized. The variational ansatz for the sc wavefunction admits a similar form as Eq.~(\ref{eqHcck}),
\begin{align}
    \ket{\Psi_{\rm sc}(\vec{k})} = L^{-1}\sum_{\vec{j}_{\rm c}} e^{i\vec{k}\cdot\vec{j}_{\rm c}} \sum_{\ell=0}^{\infty} \psi_{\ell} \sum_{\Sigma:\ell_\Sigma=\ell} \ket{\vec{j}_{\rm c},\Sigma}_{\rm sc},
\end{align}
where we sum over the position of the chargon~$\vec{j}_c$. Again, we define~$\phi_\ell = \mathcal{N}_\ell^{1/2}\psi_{\ell}$ and write the wavefunction in the string length basis:
\begin{align}
    \ket{\Psi_{\rm sc}(\vec{k})} = L^{-1}\sum_{\vec{j}_{\rm c}} e^{i\vec{k}\cdot\vec{j}_{\rm c}} \sum_{\ell=0}^{\infty} \phi_{\ell}  \ket{\ell}_{\rm sc}.
    \label{eq:VariAnsatzEllBasis}
\end{align}
Since we are working in a non-ONB, the normalization condition requires~$\sum_{\ell,\ell'}\phi^{*}_{\ell'}g^{d=2}_{\ell',\ell}\phi_{\ell} = 1$.

To derive the Gram-Schmidt parton model (GSP) in~$d=2$, we define a new set of orthonormal basis states~$\ket{\tilde{\ell}}_{\rm sc}$ with~$_{\rm sc}\langle \tilde{\ell} \ket{\tilde{\ell}}_{\rm sc} = \delta_{\tilde{\ell}',\tilde{\ell}}$ by
\begin{align}
    \ket{\ell}_{\rm sc} &= \sum_{\tilde{\ell}}\mathcal{G}^{d=2}_{\tilde{\ell},\ell}\ket{\tilde{\ell}}_{\rm sc}\\
    \ket{\ell=0}_{\rm sc} &= \ket{\tilde{\ell}=0}_{\rm sc},
\end{align}
see also Eq.~(\ref{eqGramSchmidt}) for~$d=1$.

The expression for the variational energy of the hopping Hamiltonian~$\langle \Psi_{\rm sc}(\vec{k})|\H_t\ket{\Psi_{\rm sc}(\vec{k})}$ implicitly gives us the operator in matrix form~$h^t_{\ell',\ell}$ in the string length basis~$\ket{\ell}_{\rm sc}$.
The GSP hopping Hamiltonian~$\H^{\rm GSP}_t$ for~$d=2$ can then be derived by a basis transformation~$\H^{\rm GSP}_t = \hat{\mathcal{G}}^{d=2} \hat{h}^t (\hat{\mathcal{G}}^{d=2})^{-1}$ analogously to Sec.~\ref{secApxGSPLadder}.

For the hopping Hamiltonian at~$\vec{k}=\vec{0},\vec{\pi}$, we find
\begin{equation}
    \sum_{\ell''}g_{\ell',\ell''}h^{t}_{\ell'',\ell}(\vec{k})= t_\parallel e^{i\vec{k}} \left( \tau_{\ell-1}g_{\ell',\ell-1} +  \tau_{\ell}g_{\ell',\ell+1} \right)
\end{equation}
with
\begin{equation}
    \tau_{\ell} = \begin{cases} \sqrt{z}, & \text{for}~~\ell=0 \\ \sqrt{z-1}, & \text{for}~~\ell \geq 1. \end{cases}
\end{equation}
The enhancement in the factor~$\tau_0$ arises from retracing~$\ket{\Sigma}_{\rm sc}$ on the Bethe lattice, where the origin, $\ket{\Sigma=0}_{\rm sc}$, connects to $z$ longer string states instead of~$z-1$ for all other states.
In order to exactly diagonalize the GSP Hamiltonian, we need to introduce a cut-off in the string length basis at~$\ell_{\rm max}$. The cut-off in the non-ONB yields non-hermitian contributions for the boundary states after Gram-Schmidt orthogonormalization, i.e.~$(\H_t^{\rm GSP})_{\ell,{\ell_{\rm max}}} \neq (\H_t^{\rm GSP})_{{\ell_{\rm max}}, \ell}$. Therefore, we project the GSP model onto the subspace with~$\{ \ket{\tilde{\ell}}_{\rm sc} \}_{\tilde{\ell}=0,...,\tilde{\ell}_{\rm max}-1}$ to receive a well-defined hermitian Hamiltonian.

The derivation of the magnetic interaction~$J_\perp$ involves the evaluation of overlaps that arise when first applying~$\H_{J_\perp}$ on the variational wavefunction and then projecting back onto the subspace spanned by~$\{ \ket{\ell}_{\rm sc} \}_{\ell \in \mathbb{N}_0}$.
The matrix form in the string length basis is then given by

\begin{align}
    \sum_{\ell''}g_{\ell',\ell''}h^{J_\perp}_{\ell'',\ell} &=   \mathcal{N}_{\ell}\mathcal{N}_{\ell'}^{-1/2} \left[ -\mathcal{N}_{\ell'}\ell \dfrac{z-1}{z}2^{-|\ell+\ell'|} + \sum_{1 \leq \chi \leq \ell} \left( f^{\chi}_{\ell',\ell} + F^{\chi}_{\ell',\ell} \right)\right] \\
    \text{with}~~ f^{\chi}_{\ell',\ell} &= \begin{cases} \big[ 1 - 3\Theta(\sigma-\ell') \big] 2^{-|\ell-\ell'|-1} , & \text{for}~~\ell \geq \ell' \\ & \\ (z-1)^{\ell'-\ell}2^{-|\ell-\ell'|-1} , & \text{else}  \end{cases} \\
    F^{\chi}_{\ell',\ell} &= \begin{cases}\sum\limits_{\lambda \geq 1}^{\ell'-1}(z-2)(z-1)^{\lambda-1}\big[ 1 - 3\Theta(\sigma+\lambda-\ell') \big]2^{-|\ell-\ell'|+2\lambda-1} , &  \text{for}~~\ell \geq \ell' \\ & \\ \sum\limits_{\lambda \geq 1}^{\ell-1}(z-2)(z-1)^{\lambda-1+\ell'-\ell}\big[ 1 - 3\Theta(\sigma+\lambda-\ell) \big]2^{-|\ell-\ell'|-2\lambda-1} , & \text{else}.\end{cases}
\end{align}

Here, $\Theta(x)$ is the Heaviside step function and we define $\Theta(0)=0$. 

The full GSP model in $d=2$ and for~$\vec{k}=\vec{0},\vec{\pi}$ can be calculated by transforming into the Gram-Schmidt basis
\begin{align}
    \H^{\rm GSP}(\vec{k}) = \hat{\mathcal{G}}^{d=2} [ h^t(\vec{k}) + h^{J_\perp} ] (\hat{\mathcal{G}}^{d=2})^{-1}
\end{align}
and for which we can calculate the ground-state energy by ED after projecting out the boundary at~$\ell_{\rm max}$ as discussed above. In the following, we will call~$\ell_{\rm max}$ the maximum string length after removing the boundary state. 
Moreover, we subtract the zero-hole energy, which for the $d=2$ mixD bilayer model is given by
\begin{align}
    E_{0h}=-J_\perp \sum_{\ell=0}^{\ell_{\rm max}}z(z-1)^{\ell-1}.
\end{align}
We find, as expected for the sc case, the minimum to be at~$\vec{k}=\vec{\pi}$ due to the positive sign of the hopping amplitude~$+t_\parallel$. 

The ground-state energies for the spinon-chargon and chargon-chargon case can be seen in Fig.~\ref{fig:2D} in the main text, where they are compared to DMRG calculations on a cylinder. In the tight-binding regime,~$J_\perp \gg t_\parallel$, all three approaches -- DMRG, GSP model and perturbation theory -- are in excellent agreement. In the strong coupling regime,~$t_\parallel \geq J_\perp$, the DMRG and GSP model both predict positive binding energies on the order of~$J_\perp$, however, the DMRG calculations predict even stronger binding. By comparing the spinon-chargon and chargon-chargon energies separately to the DMRG energies, we find good agreement for the sc case, which corroborates the string picture of a bound spinon and chargon. The string picture gives rise to vibration and rotational excitations in the spectrum. The variational ansatz in the string length basis Eq.~(\ref{eq:VariAnsatzEllBasis}) cannot capture the rotational excitations but the vibrational states of the string. In the spectral functions considered here, the rotational excitations carry negligible spectral weight. Similar to the chargon-chargon case in the main text Fig.~\ref{fig:spectra}c), we can thus compare the sc spectrum of the GSP model to the spectrum obtained by DMRG as shown in Fig.~\ref{fig:ARPES_2D_1hole} for~$t_\parallel/J_\perp = 2$. Due to the finite time evolution in the DMRG calculation, the spectral function has a limited resolution and we do not see individual vibrational states. However, after broadening the spectrum of the GSP model obtained by ED, they both coincide well over the full range of the spectrum. For the RIM momentum~$\vec{k}=\vec{0}$, we can see the indication of the first vibrational excitation above the ground-state.

In the chargon-chargon case, the GSP model predicts the energy at momentum~$\vec{k}=\vec{\pi}$ correctly. For the ground-state at~$\vec{k}=\vec{0}$, we find deviations of about~$10\%$ compared to the DMRG calculations. This indicates that the GSP model in the chargon-chargon case has a missing ingredient, for example loop effects or coupling to magnons, which are subject to future studies. While the overall agreement of the spinon-chargon and chargon-chargon energies already points out that the string model might be an accurate description of the polaron, on the level of the binding energies it shows larger deviations because this corresponds to the difference of the spinon-chargon and chargon-chargon energies, which becomes sensitive to errors.

%%%%%%%%%%%%%%%%%%%%%%%%%%%%%%%%%%%%%%%%%%%%%%%%%%%%%
\begin{figure}[t!]
\centering
  \includegraphics[width=0.4\linewidth]{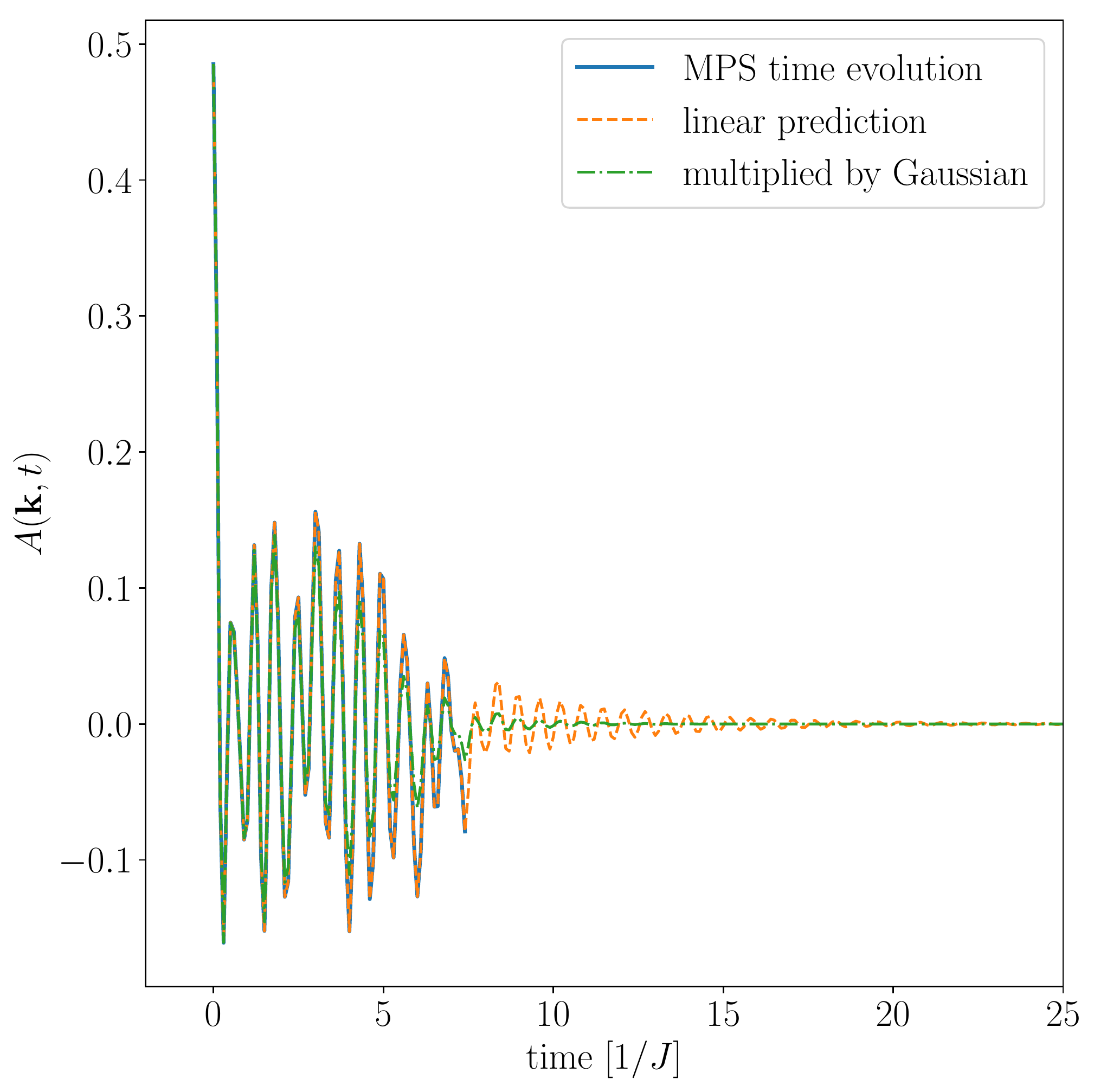}
\caption{\textbf{Time-dependence} of $A(k=0,t)$ in a mixD ladder for $t_\parallel/J_\perp = 3.4$ and $J_\parallel/J_\perp=0.31$ from time-dependent DMRG simulations with weak tunneling $t_\perp/J_\perp=0.01$ to simplify convergence. The dashed lines correspond to the linear prediction to extend the time evolution to longer times. The dashed-dotted line is the resulting signal, multiplied with a Gaussian envelope. This time-trace is then Fourier transformed to obtain the spectral function $A(k,\omega)$ shown in the main text.} 
\label{fig:spectra_time_dep}
\end{figure}
%%%%%%%%%%%%%%%%%%%%%%%%%%%%%%%%%%%%%%%%%%%%%%%%%%%%%

%%%%%%%%%%%%%%%%%%%%%%%%%%%%%%%%%%%%%%%%%%%%%%%%%%%%%
\section{DMRG simulations}
\label{secApdxDMRG}
%%%%%%%%%%%%%%%%%%%%%%%%%%%%%%%%%%%%%%%%%%%%%%%%%%%%%
Here we provide additional information about the DMRG results shown in the main text. We used the TeNPy package \cite{Hauschild2018,Hauschild2019} to perform DMRG simulations of the mixD ladder and bilayer system in the case of $N_h = 0,1,2$ holes on a system with $2 \times 40$ sites (ladder) and $2\times 12 \times 4 $ sites (bilayer). As discussed in the main text, the binding energies shown in Fig.~\ref{fig1} are obtained as $E_\text{B} = E_{2h} + E_{0h} - 2 E_{1h}$. We have carefully checked the convergence with bond dimension and obtained truncation errors on the order of $\mathcal{O}(10^{-10})$ in the case of two holes on a ladder. 

%%%%%%%%%%%%%%%%%%%%%%%%%%%%%%%%%%%%%%%%%%%%%%%%%%%%%
\begin{figure}[t!]
\centering
  \includegraphics[width=0.5\linewidth]{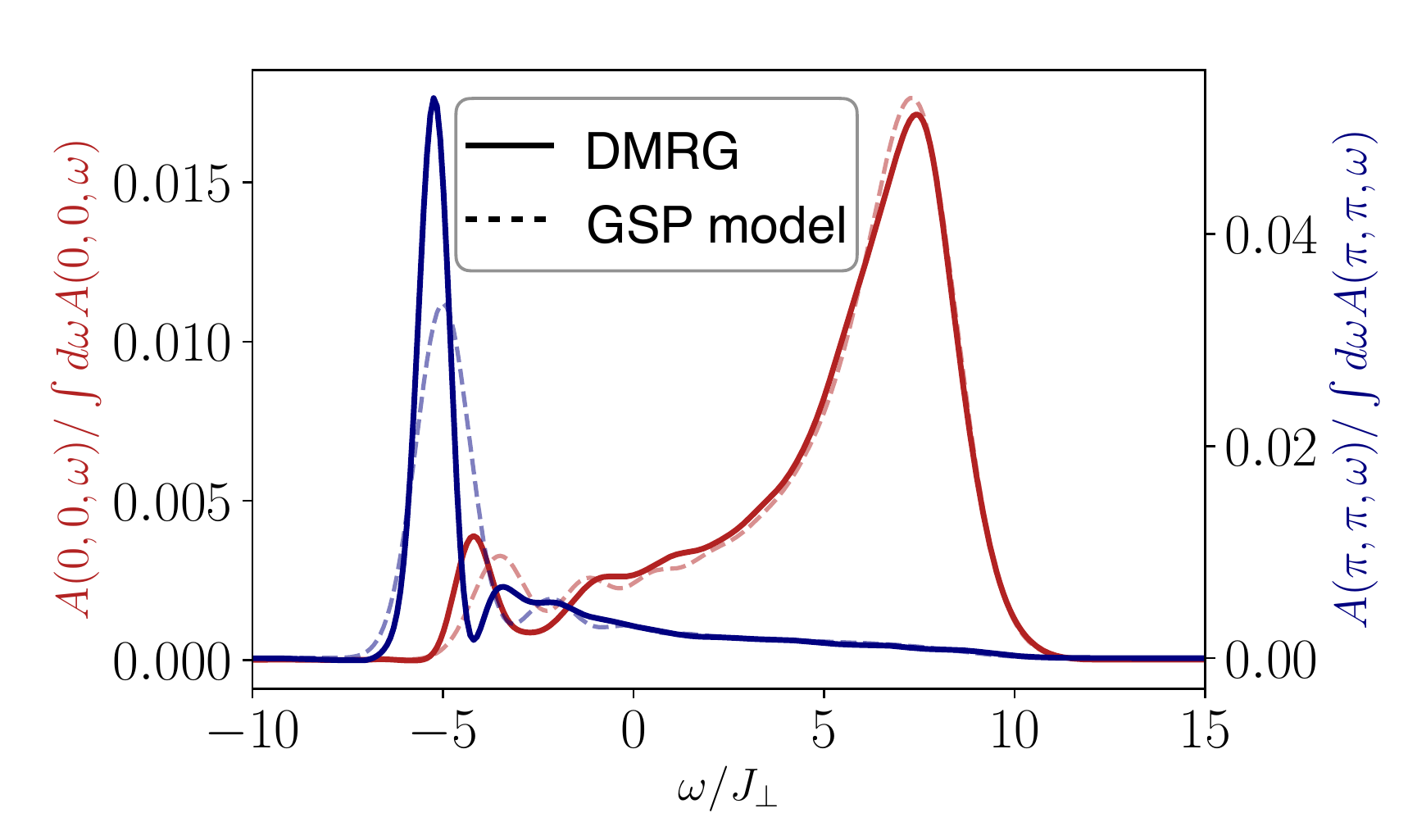}
\caption{\textbf{Spectra in 2D for a single hole} at momenta $\mathbf{k}=(0,0)$ and $\mathbf{k}=(\pi,\pi)$ calculated with DMRG (straight lines) and from the GSP model (dashed lines). For the GSP model, a Gaussian function with $\sigma_0 = 0.5$ is used to broaden the delta peaks.} 
\label{fig:ARPES_2D_1hole}
\end{figure}
%%%%%%%%%%%%%%%%%%%%%%%%%%%%%%%%%%%%%%%%%%%%%%%%%%%%%

In order to calculate the one- and two-hole spectra shown in Fig.~\ref{fig:spectra} of the main text and Figs.~\ref{fig:quadratic_fit} and \ref{fig:ARPES_2D_1hole} in the supplement, we start by calculating the ground state $\ket{\psi_0}$ without a hole. We then apply the operator $\hat{C}_{(1/2),i=L/2}$ as defined in the main text, where $L=40$ is the length of the ladder (cylinder). The new state is then time-evolved using the $W^{(II)}$ method introduced in \cite{Zaletel2015}. By calculating the overlap with $\hat{C}_{(1/2),j}\ket{\psi_0}$ the time-dependent correlation function $C_{(1/2),ij}(t)$ is obtained up to a finite time $t_0$. After Fourier transforming to momentum space, we use linear prediction to increase the time window and multiply the resulting data with a Gaussian envelope $w(t) = \exp\left[ -0.5 (t\sigma_\omega)^2\right]$ with $\sigma_\omega=1.5/t_0=0.2J$ in order to minimize the weight of the data generated by linear prediction on the resulting spectral function. The time-traces for the different steps are exemplary shown for the two-hole spectra in $d=1$ at $k=0$ in Fig.~\ref{fig:spectra_time_dep}.

%%%%%%%%%%%%%%%%%%%%%%%%%%%%%%%%%%%%%%%%%%%%%%%%%%%%%
\begin{figure}[t!]
\centering
  \includegraphics[width=0.5\linewidth]{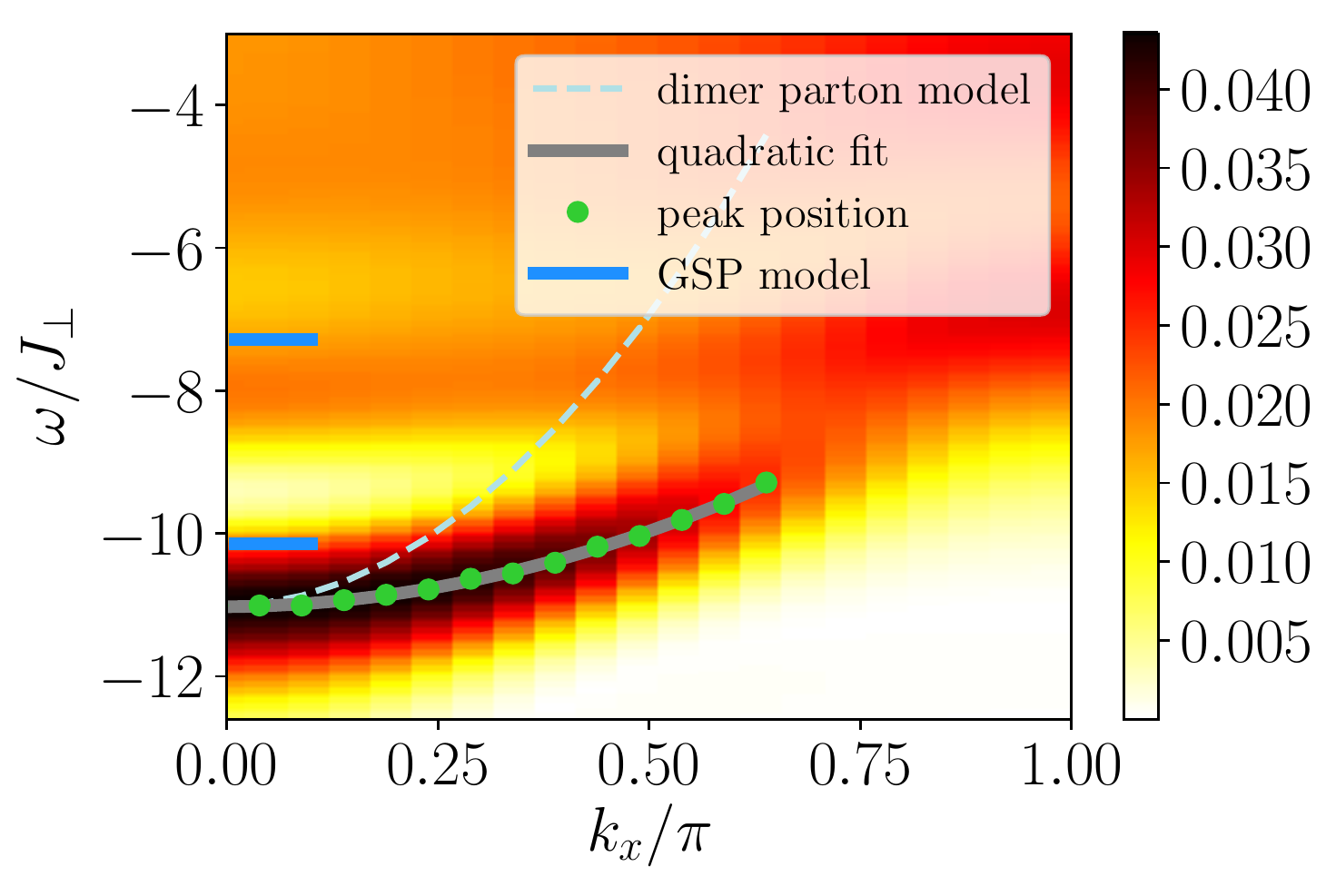}
\caption{\textbf{Spectra in 2D around $\mathbf{k}=(0,0)$} for two holes at low energies. Green dots denote the extracted position of the lowest energy peak. Gray line corresponds to a quadratic fit to the extracted peak positions. The blue lines show the ground state energy and first vibrational excitation predicted by the GSP model. } 
\label{fig:quadratic_fit}
\end{figure}
%%%%%%%%%%%%%%%%%%%%%%%%%%%%%%%%%%%%%%%%%%%%%%%%%%%%%

The effective mass of the pair in $d=2$ is determined from a quadratic fit around $\mathbf{k}=(0,0)$ to the position of the lowest peak. The extracted peak positions, as well as the quadratic fit are shown in Fig.~\ref{fig:quadratic_fit}.

\end{document}